\documentstyle[psfig]{ioplppt}
\newcommand{\bma}{\begin{displaymath}} \newcommand{\ema}{\end{displaymath}}
\newcommand{\bit}{\begin{itemize}} \newcommand{\eit}{\end{itemize}}
\newcommand{\beq}{\begin{equation}} \newcommand{\eeq}{\end{equation}}
\newcommand{\bce}{\begin{center}} \newcommand{\ece}{\end{center}}
\newcommand{\eqref}[1]{~(\ref{#1})}

 \newcommand\ie{\it i.e.\/\rm,\ }
\newcommand\eg{\it e.g.\/\rm,\ }

\newcommand{\ltsima}{$\; \buildrel < \over \sim \;$} \newcommand{\gtsima}{$\;
\buildrel > \over \sim \;$}

\newcommand{\simlt}{\lower.5ex\hbox{\ltsima}}
\newcommand{\simgt}{\lower.5ex\hbox{\gtsima}}

\newcommand{\loeq}{\mathrel{\hbox{\lower1ex\hbox{\rlap{$=$}\raise1.2ex
\hbox{$<$}}}}}
\newcommand{\goeq}{\mathrel{\hbox{\lower1ex\hbox{\rlap{$=$}\raise1.2ex
\hbox{$>$}}}}}

\newcommand{\la}{\mathrel{\mathchoice {\vcenter{\offinterlineskip\halign{\hfil
$\displaystyle##$\hfil\cr<\cr\sim\cr}}}
{\vcenter{\offinterlineskip\halign{\hfil$\textstyle##$\hfil\cr<\cr\sim\cr}}}
{\vcenter{\offinterlineskip\halign{\hfil$\scriptstyle##$\hfil\cr<\cr\sim\cr}}}
{\vcenter{\offinterlineskip\halign{\hfil$\scriptscriptstyle##$\hfil
\cr<\cr\sim\cr}}}}}

\renewcommand{\br}{{\mathbf r}} \newcommand{\bomega}{{\mathbf \omega}}
\newcommand{\bx}{{\mathbf x}} \newcommand{\bv}{{\mathbf v}}
\newcommand{\bk}{{\mathbf k}} \newcommand{\bq}{{\mathbf q}}
\newcommand{\bM}{{\mathbf M}} 
\newcommand{\bp}{{\mathbf p}}

\newcommand{\VEV}[1]{\left\langle #1\right\rangle}
\newcommand{\partder}[2]{{\partial #1\over\partial #2}}
\newcommand{\hub}{\frac{ \dot{a} }{a} } \newcommand{\dt}{{\partial_t}}

\newcommand{\rx}{{\rm x}} \newcommand{\rmq}{{\rm q}}

\newcommand{\cD}{{\cal D}} \newcommand{\cI}{{\cal I}} \newcommand{\cT}{{\cal
T}}

\newcommand{\bxp}{{\bx'}}

\renewcommand{\bar}{\overline} \newcommand{\eps}[1]{\varepsilon^{#1}}
\newcommand{\ord}[1]{{\cal O(\varepsilon^{\rm #1})}}
\newcommand{\del}[1]{\delta^{(#1)}} 
\newcommand{\dis}[1]{\Psi^{(#1)}} 
\newcommand{\bdis}[1]{\bPsi^{(#1)}}
\newcommand{\bdist}[1]{\tilde{\bPsi}^{(#1)}}
\newcommand{\dist}[1]{\tilde{\Psi}^{(#1)}}

 \newcommand{\jac}[1]{J^{(#1)}}
\newcommand{\jjac}[1]{J^{(#1)\; 2}} \newcommand{\jjjac}[1]{J^{(#1)\;3}}
\newcommand{\jjjjac}[1]{J^{(#1)\; 4}} \newcommand{\Var}{\sigma_1^2}
\newcommand{\VVar}{\sigma_1^4}

\newcommand{\K}[1]{K^{(#1)}} 
\renewcommand{\L}[1]{L^{(#1)}} \newcommand{\M}[1]{M^{(#1)}}

\newcommand{\VVVar}{\sigma^6} \newcommand{\Skew}{S} \newcommand{\Kurt}{K}

\newcommand{\iras}{{\sl IRAS}}
\psnoisy 

\begin{document}

\title{INTRODUCTORY OVERVIEW OF EULERIAN \& LAGRANGIAN PERTURBATION
THEORIES}[OVERVIEW OF PERTURBATION THEORIES]

\author{F. R. BOUCHET} \address{Institut d'Astrophysique de Paris, CNRS, 98bis
Boulevard Arago, F--75014 Paris France.}

\tableofcontents

\bigskip

These lectures notes give an introduction to the fast developing area of
research dealing with perturbative descriptions of the gravitational
instability in an expanding universe. I just sketch the outlines of some
proofs, and many important contributions (and references) were left out. Many
untouched aspects are reviewed in \cite{sah_col} which also contains a useful
reference list.

In the notes dedicated to the Eulerian approach (\S\ref{sec:fluid} -
\S\ref{sec:further}), I derive the fluid equations starting from a microscopic
description, I give their linear solution and some higher order corrections,
and describe a number of applications concerning the evolution of the one point
probability density function for the density contrast field and for the
divergence of the corresponding velocity field.

In the notes dedicated to the Lagrangian point of view (\S\ref{sec:zeldo} -
\S\ref{sec:lag_app}), I recall Zeldovich Approximation and its higher orders
corrections, and present some applications, like their use in describing the
statistical effect of the real space-redshift mapping, or as approximate
description of the full non-linear dynamics.


\section{THE FLUID EQUATIONS\label{sec:fluid}}

In this section, I derive the fluid equations as velocity moments of the full
microscopic evolution equations taken in the mean field, or continuum, limit.
We shall restrain to the case of perturbations which may be described in the
Newtonian approximation.

In all the following, we use comoving coordinates $\br = a \bx$, and $\bp = m
a^2 \dt\bx$, where $m$ stands for the particles mass, and $a$ for the metrics
scale or expansion factor. In that case, the motion and field (Poisson)
equations write
\[
        \dt\bp = - \nabla \phi,\quad \nabla^2 \phi = 4\pi G a^2 \bar\rho
        \delta,
\] 
if $\delta\equiv \rho/\bar\rho -1$ stands for the density contrast field. The
solution to Poisson equation is
$ \phi(x)=-Ga^2 \int d\bx\prime (\rho(\bxp) -\rho_b) / |\bxp - \bx|$, where
$\rho(\bxp)$ fluctuates in an homogeneous and isotropic way around the
background value $\rho_b$. Thus
\[
        -\nabla \phi = Ga^2 \int (\rho(\bxp) -\rho_b) \frac{\bxp-\bx}{|\bxp -
        \bx|} d\bx\prime = Ga^2 \int \rho(\bxp) \frac{\bxp-\bx}{|\bxp - \bx|}
        d\bx\prime
\] 
where it is implicit that, in the last integral, one follows the prescription
of integrating first over angle and only then over $\bxp$.

\subsection{Microscopic Description}

Let us denote by $f(\bx, \bp, t)$ the one-particle distribution function,
i.e. the probability density of finding at time $t$ a {\em collisionless}
particle within the infinitesimal phase space volume $d\mu=d\bx d\bp$.  For
$N_p$ point particles, each with a trajectory given by $\bx_i(t)$ and
$\bp_i(t)$, the system is fully described by
\[
        f(\bx, \bp, t)= \sum_{i=1}^{N_p} \delta_D(\bx - \bx_i(t) ) \delta_D(\bp
                        - \bp_i(t) ),
\] 
where $\delta_D$ stands for Dirac's function.  Particle conservation plus
Liouville theorem (i.e. $\frac{d\mu}{dt} = 0$) give the equation governing the
evolution of the system, $\frac{df}{dt} = 0$.

In our discrete particle model, the density is
\[
        \rho = \rho_b [ 1+ \delta(\bx, t)] = \frac{m}{a^3} \int f(\bx, \bp, t)
        \bp = \frac{m}{a^3} \sum_i \delta_D (\bx - \bx_i) ,
\] 
and one recovers the familiar form $ -\nabla \phi = Ga^2 \sum_i \frac{\bx_i -
\bx}{|\bx_i - \bx|^3}$. By using the motion equation, we then get the {\em
Klimontovich equation}
\[
        {\rm d}_t f = \dt f + \frac{\bp}{m a^2}\cdot\nabla_x f - m \nabla \phi
        \cdot \nabla_p f =0 ,
\] 
a complete description with the field equation
\[
        -m \nabla \phi = \frac{Ga^2}{a} \int f(\bxp, \bp', t) \frac{\bx\prime -
        \bx}{|\bxp - \bx|^3} d\bxp d\bp' .
\]

\subsection{Macroscopic Description}

Let us now consider a smoothed out (phase space) density obtained by averaging
over patches of size intermediate between the microscopic level and the system
size, so that we can take ``local'' averages. In this ``coarse-grain''
distribution, we loose the discrete character of the particles. This can also
be seen as a continuum (or thermodynamical) limit such that one formally takes
the limit of an infinite number of particle, $N_p \longrightarrow \infty$,
while their individual mass goes to zero, $m \longrightarrow 0$, keeping the
mass density $N_p m$ constant.

In this {\em mean field} limit, the Klimontovich equation becomes the {\em
Vlasov equation}. It has formally the same writing, but now the solutions $f$
are to be found in the ensemble of smooth, continuous, functions (since the
discrete character embodied by Dirac function has been lost). Note that by
loosing the discrete character of the particles, one of course cannot describe
the effect of direct encounters (e.g. two-relaxation effects). In other words,
one only considers the evolution of perturbations through {\em fluctuations of
the mean field} created collectively by many particles.

Intuitively, this collective description of the evolution should be legitimate
when the 2-body potential energy at the mean interparticle distance $n^{-1/3}$,
$G m^2 n^{1/3}$, is much smaller than their average kinetic energy, $ m
\sigma_v^2$, which is indeed confirmed by more rigorous analysis.

\subsection{Velocity Moments and the Fluid Limit}

A classical method amounts to find equations between various velocity
(momentum) moments of the one-point distribution function $f(\bx, \bp,
t)$. These are simply obtained by multiplying the equation governing the
evolution of $f$ (here the Vlasov equation) by products of velocity
components(or here momentum components $p^\alpha$) before integrating over all
the velocities components.

\underline{The zeroth moment} relates the density $\rho(x) = (m/a^3) \int \bp
f(\bx, \bp, t) $ to the mean velocity of a phase space element,
$ \bv = \VEV{\bp/ma} =\int (\bp/ma) f\, d\bp / \int f\, d\bp $ by the {\em
continuity equation}
\begin{equation}
        \dt \rho + 3 \hub \rho + \frac{1}{a} \nabla\cdot \rho\bv = 0 ,
        \label{mom0}
\end{equation}  
 which is often rewritten as $\quad \dt \delta + \frac{1}{a} \nabla \left[
(1+\delta) \bv \right] = 0$.

\underline{The first moment} obtains by multiplying Vlasov equation by the
$\alpha$ component of momentum, $p^\alpha$, which yields the velocity equation
\begin{equation}
        \dt \int p^\alpha d\bp + \frac{1}{a^2} \partial_\beta \int p^\alpha
        p^\beta f d\bp + a^3 \rho \partial_\alpha \phi = 0 .
\label{mom1}
\end{equation} 
It is then convenient to introduce the {\em centered} second moment tensor,
i.e. the ``pressure tensor'', $P^{\alpha\beta}$, defined by 
\[
        \VEV{\frac{p^\alpha}{ma} \frac{p^\beta}{ma}} = \frac{ P^{\alpha\beta}
        }{\rho} + v^\alpha v^\beta ,
\]
to rewrite the source term in the mean velocity equation\eqref{mom1}.

\underline{Truncating the hierarchy:} the second moment equation would in turn
relate the pressure tensor to a ``heat flux tensor'' $\VEV{p^\alpha p^\beta
p^\gamma}$. Of course, one obtains by this procedure a hierarchy of equations,
each relating the evolution of a moment to the one of next higher order.

Truncating the hierarchy amounts to impose a relation between a given moment
and the lower ones. In particular, one often assumes an isotropic pressure
tensor
\[ 
\frac{P^{\alpha\beta} }{\rho} = \delta_{\alpha\beta} \frac{p}{\rho},
\]
where $\delta_{\alpha\beta}$ is Kronecker's symbol, and $p$ is the usual
pressure. In that case the first moment equation\eqref{mom1} becomes the {\em
Euler equation}
\[
        \dt v^\alpha + \hub v^\alpha + \frac{(\bv \cdot \nabla) v^\alpha}{a} =
        - \frac{\partial_\alpha p}{a \rho} - \frac{\partial_\alpha \phi}{a} ,
\] 
which together with the continuity equation and an equation of state relating
the pressure and the density forms a closed system of equation, often referred
to as the ``fluid equations''.

Note though that till now we have said nothing about the role of
``collisions''. And indeed, particles interacting only gravitationally like
neutrinos should have an isotropic velocity dispersion everywhere, as long as
one does not reach the multi-stream regime, i.e. before ``shell crossing''. In
fact, early on, the velocity dispersion can often be neglected altogether,
$p=0$, and the equations above describe a so-called ``dust universe''.  But of
course, one expects on very general grounds (see the Lagrangian section for
more details) to have an anisotropic, ``pancake-like'' collapse, which cannot
be followed any more by these equations (although one may attempts to guess an
``effective'' equation of state mimicking macroscopically the complex
microscopic dynamics).

On the other hand, if we have a {\em bona fide} fluid made of baryons, their
collective interactions do maintain a local isotropy and establish an equation
of state, with e.g. $p(\rho) = p(\rho_b)+c_s^2 \rho_b \delta$, where $c_s$
stands for the sound speed, $c_s = ( dp/d\rho )^{1/2}$. In any case, the point
is that the ``fluid equations'' can properly be used to describe (some of) the
evolution of dark matter collisionless particles.

\underline{Other forms of the continuity and Euler equations:} it is often
convenient to introduce the conformal time $\tau$ defined by $d\tau =
dt/a$. Denoting the partial derivative versus $\tau$ by an overdot, and
redefining $\bv$ as $\dot{\bx} $, the previous equations become
\begin{equation}
        \dot{\delta} + \nabla \left[ (1+\delta ) \bv \right] = 0 \quad {\rm
        and}\quad \dot{\bv} +\hub \bv + (\bv \cdot \nabla) \bv = - \frac{\nabla
        p}{\rho} - \nabla \phi .
\label{eq:euler_conform}
\end{equation} 
Finally, it is advantageous for some applications to rather introduce $D(t)$ as
the time coordinate, $D(t)$ standing for the linear growth rate of density
fluctuations (see below).

\section{EULERIAN PERTURBATIVE SOLUTIONS}

\subsection{Linear Density Field}

By inserting the first moment equation\eqref{mom1} into the continuity
equation\eqref{mom0} one gets
\begin{equation}
%
%
        \partial_{t}^2 \delta + 2 \hub \dt \delta = \frac{1}{a^2} \nabla \left[
        (1+\delta ) \nabla \phi \right] + \frac{1}{a^2} \partial_\alpha
        \partial_\beta \left[ \frac{P^{\alpha \beta}}{\rho_b} + (1 + \delta)
        v^\alpha v^\beta \right] .
\label{eq:dens}
\end{equation} 
The Eulerian perturbative regime is when
\[      
        \delta \ll 1 \qquad {\rm and} \qquad \left( \frac{v t}{d} \right)^2
        \sim \delta \ll 1 ,
\] 
where $d$ stands for the coherence length of the perturbation and $t$ its
dynamical time, $t \sim (G \rho)^{-1/2}$ (in an $\Omega =1$ universe, the FRW
equations lead to $6 \pi G \rho t^2 =1$ in the matter era). The second
requirement is simply that gradients should be small. In {\em linear}
perturbation theory, one neglects all term in $\delta^2, v^2, \delta v$, and
the fluid equations assume the form
\begin{equation}
        \dt \delta + {\nabla \cdot \bv}{a} = 0 \qquad {\rm and} \qquad
        \partial_{t}^2 \delta + 2 \hub \dt \delta = \frac{\nabla^2 p}{\rho_b
        a^2} + 4 \pi G \rho_b \delta .
\label{eq:lin}
\end{equation} 
Solutions are known for many different cases, i.e. for different equation of
state, and different time evolution of the factor of expansion $a$. Here we
recall some of the most useful ones.

\subsubsection{Dust Universe, $p=0$:}

since there are now no spatial derivatives in the equations, the evolution is
self-similar for arbitrary density contrasts $\delta$. Let us denote by $D_a$
and $D_b$ the two growth rates solution of the second order differential
equation, and by $A$ and $B$ the associated spatial parts, so that the solution
if the linear combination
\[
        \delta(\bx, t) = A(\bx) D_a(t) + B(\bx) D_b(t) ,
\] 
$D_a$ corresponding by convention to the fastest growing mode

\underline{In the Einstein-de Sitter case,} $\Omega =1$ and $a \propto
t^{2/3}$. Thus eq.~(\ref{eq:dens}) becomes $\partial_{t}^2 \delta +
\frac{4}{3t} \dt \delta = \frac{2}{3 t^2} \delta$, and we have $D_a = t^{2/3}$
and $D_b = t^{-1}$. Note that the exponential growth of a static universe (when
$\dot a =0$) is now much slower.

\underline{In the open case,} full solution may be obtained by changing the
time coordinate to $a$ given by
\[
        \left( \hub \right)^2 = \frac{8 \pi G \rho_b}{3} \left[ 1 +
        (\Omega_0^{-1}-1) \left( \frac{a}{a_0} \right) \right] .
\] 
But in the asymptotically late regime when the curvature term dominates,
i.e. $a \propto t$, one simply has $\partial_{t}^2 \delta + \frac{2}{t} \dt
\delta + \frac{3 \Omega}{2 t^2} \delta \sim \partial_{t}^2 \delta + \frac{2}{t}
\dt \delta = 0$, since $4 \pi G \rho_b \ll t^{-2}$. Thus $D_a = 1$ and $D_b =
t^{-1}$; At best the density contrast is frozen in the free expansion
regime. And this happens when $(\Omega_0^{-1}-1)a/a_0 \sim 1$, i.e when $1+
z_{freeze} = \Omega_0^{-1}-1$. In other words, a perturbation grows like
$t^{2/3}$ at $z \gg z_{freeze}$ and stops afterwards.

\underline{Fixing $A(\bx)$ and $B(\bx)$:} the solution to our second order
differential equation is fully specified by giving, e.g. the initial density
contrast $\delta_i$ and velocity $\bv_i$ at time $t_i$,
\[
        \delta_i = A D_a(t_i) + B D_b(t_i), \quad{\rm and}\quad - \frac{\nabla
        \cdot \bv_i}{a_i} = A \left. \frac{d D_a}{d t} \right|_i + B
        \left. \frac{d D_b}{d t} \right|_i .
\] 
Introducing the Wronskian $W = D_a(t_i) \left. \frac{d D_b}{d t}\right|_i -
D_b(t_i) \left. \frac{d D_a}{d t}\right|_i$, we have then
\[
        \delta = \frac{\delta_i}{W} \left[ D_a \left. \frac{d D_b}{d
        t}\right|_i - D_b \left. \frac{d D_a}{d t}\right|_i \right] +
        \frac{\nabla \cdot \bv_i}{a_i} \left[ D_a D_b(t_i) - D_b D_a(t_i)
        \right] .
\] 
In the Einstein-de Sitter case, $W = - \frac{5}{3} \frac{D_a(t_i)
D_b(t_i)}{t_i}$. If the initial velocity may be neglected, one then finds
\[
        \delta = \delta_i \left[ \frac{3}{5} D_a + \frac{2}{5} D_b \right] ,
\] 
with $\delta = \frac{3}{5} D_a \delta_i$ at late times (in the following, we
shall drop the subscript $a$, and $D$ will refer to the fastest growth
rate). On the other hand, in numerical simulations, one usually sets the
velocity field to select the growing mode only.

\subsubsection{Jeans length of an ideal fluid with $p=p(\rho)$:}

plugging $p(\rho) = p(\rho_b)+c_s^2 \rho_b \delta$ in\eqref{eq:lin}, we get
\[
        \partial_{t}^2 \delta + 2 \hub \dt \delta = \left( \frac{c_s}{a}
        \right)^2 \nabla^2 \delta + 4 \pi G \rho_b \delta .
\] 
It's Fourier transform is
\[
        \partial_{t}^2 \delta_k + 2 \hub \dt \delta_k = \left[ 4 \pi G \rho_b -
        \frac{c_s^2 k^2}{a^2} \right]^2 \delta_k ,
\] 
which shows that there is a characteristic wavelength $\lambda_J = 2 \pi / k_J
= \pi \frac{c_s}{G \rho} \sim c_s t$ for the scale of a perturbation. Above
this Jeans length, the pressure cannot counteract the self-gravity of the
perturbation ($c_s t < \lambda$), while at $\lambda < \lambda_J$ the
perturbation oscillates like a sound wave.

\subsection{Linear Peculiar Velocity Field}

\subsubsection{Solution in a dust Universe:}

the linearized continuity and velocity equation in the pressureless case reduce
to
\[
        \dt \delta + \frac{1}{a} \nabla \bv = 0 \quad {\rm and}\quad \dt \bv +
        \hub \bv = - \frac{\nabla \phi}{a} .
\] 
Since $\delta \propto D$, we have $\nabla\cdot\bv = a \partial_t \delta = - Ha
f \delta = -\frac{H a f}{4 \pi G \rho_b a^2} \nabla^2\phi$ where $H$ stands for
Hubble's ``constant'', $\dot{a}/a$, and $f(\Omega) = (a/D)\, dD/da$ can be well
fitted by $\Omega^{0.6}$ over a large range. Thus
\[
        \bv = \frac{2}{3} \frac{f}{H\Omega} \frac{-\nabla\phi}{a} +
        \frac{\nabla \times U(\bx)}{a} ,
\] 
which satisfies the first equation for an arbitrary $U(\bx)$. This last term
decays as $a^{-1}$, as does $v$ in the absence of a driving term. Since it
quickly becomes negligible, we find that at the linear order {\em the velocity
is parallel to the acceleration} at late times, the proportionality constant
depending rather strongly on the density parameter.

\subsubsection{Application to data:}

the strong $\Omega$ dependence of $f(\Omega)$ has been used in the context of
the linear bias model, in which one assumes that, at least on sufficiently
large scales, the density contrast field computed from galaxies, $\delta_g$, is
simply on average proportional to the underlying density field, i.e. $\delta_g
= b \delta$ (although that may turn out to be a poor approximation for some
applications). In that case, the velocity field is proportional to the
parameter $\beta = f(\Omega)/b$. Since one may estimate $\delta_g$ through
\def\cPhi{\bPhi}
\[
        1+\delta_g = \frac{1}{n} \sum_i \frac{1}{\cPhi(\bx_i)}\delta_D(\bx -
        \bx_i) \quad{\rm with}\quad n = \frac{1}{V} \sum_i
        \frac{1}{\cPhi(\bx_i)} ,
\]
where the density $n = \frac{1}{V} \sum_i \frac{1}{\cPhi(\bx_i)}$ obtains from
the catalog selection function $\cPhi$, one may then compare the predicted
velocity from the observed galaxy density field, $\bv_{pred}$ to the
``observed'' one. One thus try to find the $\beta$ which best match the two
velocity fields, with
\[
        \bv_{pred} = \frac{\beta H a}{4 \pi} \int_{x < x_{max}} \delta_g (\bx')
        W(|\bx' - \bx|) \frac{\bx'-\bx}{|\bx' - \bx|}\, d^3x' .
\]
Note that in practice, we do not have an infinite catalog, and we have to
truncate the integral at a maximal radius $x_{max}$ selected such that the
signal still dominates over the shot noise (i.e. beware of $1/\cPhi$ !). Even
worse, not all the sky is surveyed homogeneously and isotropically (e.g. close
to the galactic plane), so that the density field has to be extrapolated in
some way. Note also the introduction of a window function $W$ (whose size is
typically greater than 500 km/s) to average over a rather large volume in order
to reduce the level of non-linearities, reduce the shot noise, the distance
uncertainties and the impact of the triple-value zone in relating redshifts to
distances ($\dot{a} x =c z - v$ is solved iteratively)

Alternatively, one could use the proper velocities of galaxies to compare with
the observed density field. This area has been the subject of intense research
in the last few year, and will be covered in much greater detail in the notes
of A. Dekel (this volume). For a more in-depth view, one could look at the
proceedings~\cite{cvf} of the ``Cosmic Velocity Field'' meeting held in 1993,
or the recent review~\cite{velrev}.

\subsection{Non-Linear Corrections\label{sec:ewnl}}

Let us now come back to the density field equation\eqref{eq:dens} for a
pressureless medium,
\begin{equation}
        \partial_{t}^2 \delta + 2 \hub \dt \delta = \frac{1}{a^2} \nabla \left[
        (1+\delta ) \nabla \phi \right] + \frac{1}{a^2} \partial_\alpha
        \partial_\beta \left[ (1 + \delta) v^\alpha v^\beta \right] .
\label{eq:dens_dust}
\end{equation}
In order to find a description of the evolution which may remain valid longer
than the linear one, we write the density contrast and the velocity fields as
perturbative expansions
\[
        \delta (\bx, t) = \delta^{(1)}(\bx, t) + \delta^{(2)}(\bx, t) +
        \ldots\, ,
\]
\[
        \bv (\bx, t) = \bv^{(1)}(\bx, t) + \bv^{(2)}(\bx, t) + \ldots\, ,
\]
where $\delta^{(1)}(\bx, t)$ and $\bv^{(1)}(\bx, t)$ are the linear terms
obtained previously. They are of order $\delta_i \equiv \varepsilon$ (if we
write $\delta^{(1)}(\bx, t) = \varepsilon(\bx) D(t)$, $\bv^{(1)\alpha}(\bx, t))
=(f/H) \partial_\alpha \Phi_i$, if $\Phi_i$ stands for the initial potential
$\int \frac{d^3x}{4 \pi} \frac{\varepsilon(\bx')}{|\bx-\bx'|}$, and we retain
only the fastest growing modes). The next term $\delta^{(2)}= O(\delta^{(1)})^2
= O(\varepsilon^2)$ is a non-linear correction, solution of the equation when
only quadratic non-linearities are retained. The source terms are then products
of the linear solutions. The resulting second order solution can then be used
iteratively to obtain still higher order solutions...

The solution to the equation of order two in $\varepsilon$ is
\begin{equation}
        \delta^{(2)} = D(t)^2 \left[ \frac{2}{3} \left[ 1 + \kappa(t) \right]
        \varepsilon^2(\bx) + \nabla \varepsilon(\bx) \cdot \nabla \Delta +
        \left[ \frac{1}{2} - \kappa(t)\right] \cT^2 \right]\ ,
\label{e2_solution}
\end{equation}
where $\cT^2 = \sum_{\alpha,\beta = 1}^3 (\cT_{\alpha\beta})^2$ is the
contraction of the (linear order) tide tensor
\begin{equation}
        \cT_{\alpha\beta}= \nabla_{\alpha} \nabla_{\beta} \Phi_i -\frac{1}{3}
        \delta_{\alpha\beta}\nabla^2 \Phi_i
\label{def:tide_tens}
\end{equation}
whose presence can only speed up the collapse. The parameter $\kappa (t)$ is a
slowly varying function of cosmological time which is well approximated
(Bouchet \etal 1992) by
\begin{equation}
        \kappa \approx {3\over 14}\Omega^{-2/63} ,
\label{kappa-omega}
\end{equation}
in the range $0.05 \leq \Omega \leq 3$ (the accuracy of this approximation is
then better than 0.4\%).  For $\Omega = 0$, $\kappa = {1\over4}$. The exact
expression for $\kappa (\Omega)$, valid in the entire range $\Omega \geq 0$, is
given in~\cite{omega}).  For $\Omega =1$, $\kappa ={3\over 14}$, and we find
\[
      \delta = \delta^{(1)} + \frac{5}{7} \delta^{(1)2} - \partial_\alpha
      \delta^{(1)} \partial_\alpha \Phi_i + \frac{2}{7} \partial_{\alpha \beta}
      \Phi_i \partial_{\alpha \beta} \Phi_i ,
\]
the classical Einstein-de Sitter solution (\eg \cite{lss}, hereafter LSS,
eq. [18.8]).

\paragraph{Vorticity and Shear}

By taking the curl of Euler equation\eqref{eq:euler_conform} when the conformal
time $\tau$ is used, we find that gravity alone cannot generate
vorticity. Indeed, introducing $\bomega = \nabla \times \bv$, we get (an
overdot denoting partial derivatives versus $\tau$)
\[
        \dot{\bomega} = \nabla \times (\bv \times \bomega) - \hub \bomega
        \frac{\nabla\rho \times \nabla p}{\rho^2} .
\]
Thus $\dot{\bomega} = 0$ for a pressureless fluid which is initially
vorticity-free. Furthermore, the density field equation\eqref{eq:dens_dust} for
a pressureless medium can be cast in the following form
\[
        \ddot{\delta} + \hub \dot{\delta} = \frac{4}{3} \frac{\dot{\delta}}{1 +
        \delta} + (1+\delta ) \left[ \sigma_{\alpha\beta}\sigma^{\alpha\beta} -
        \frac{1}{2} \omega^2 + 4 \pi G \rho_b \delta \right] ,
\]
where the velocity shear tensor now has the form $\sigma^{\alpha\beta} =
\frac{1}{2} \left[ \frac{\partial \dot{\bx}^\alpha}{\partial x^\beta} +
\frac{\partial \dot{\bx}^\beta}{\partial x^\alpha} \right]$. Thus the shear
accelerates the collapse, while vorticity slows it down. This is this
acceleration, absent at the linear order, that showed up at the second order
[eq.\eqref{e2_solution}] through the (linear) tide tensor ($\cT^{\alpha\beta} =
\frac{H f D}{4 \pi} \sigma^{\alpha\beta} - \frac{ \delta_{\alpha\beta}
}{3}\frac{ \nabla \cdot \bv}{a}$).  Finally, let us stress that the rate of
collapse depends on the three eigenvalues of the tide tensor, which implies
that the highest peaks of the density field do not compulsorily collapse first
(see \S\ref{sec:pancakes}).

\section{SOME BASIC APPLICATIONS}

\subsection{The PDF as a Measure of Large Scale Structures}

Many approaches can be used to characterize statistically the observed large
scale structures in the Universe. Here we focus on the one-point distribution
function, or PDF. Empirically, this approach was first applied by Hubble in
1934 \cite{hub}. Let $Q(r)$ be a physical quantity whose value depends on the
location. Let $Q_\ell(r)$ be the smoothed value of $Q$ on scale $\ell$, \ie
$Q_\ell(r)$ is obtained by convolving the scalar field $Q(r)$ by a smoothing
window $W_\ell$ of characteristic scale $\ell$ ($Q_\ell(\br) = Q({\bf
s})*W_\ell(\br-{\bf s}) = \int d^3s\, Q(s)\, W_\ell(r-s)$). The probability
that $Q_\ell$ lies between $Q$ and $Q+dQ$ is $P(Q, \ell)\; dQ$, $P$ being its
PDF.

If $Q \equiv N$ where $N$ is a number of galaxies, and $W$ is a top-hat window,
then the PDF gives the probability of finding $N$ galaxies in a sphere (in 3D)
of size $\ell$.  In the following, $Q$ will be either the mass density contrast
field $\delta =\rho/\bar{\rho}-1$, or the divergence of the associated velocity
field in units of the Hubble constant, $\theta = \nabla \cdot \bv / H$; we will
consider the two most commonly used types of windows, a top-hat and a gaussian
one. Later on, we shall discuss the relation of the mass distribution to the
distribution of luminous galaxies.

It is frequently assumed that, very early in the history of the Universe, the
density contrast $\varepsilon \equiv \delta (t= t_i)$ could be taken as
normally distributed, \ie
\[
        P(\varepsilon, \ell) \propto \exp\left[ -{\varepsilon_\ell^2 \over 2\,
        \sigma_i^2(\ell)} \right],
\]
where $\sigma_i^2(\ell) \equiv \VEV{\varepsilon_\ell^2}$ stands for the initial
variance of the density contrast field smoothed on scale $\ell$. It is assumed
to be small at all scales, $\sigma_i \ll 1$. The second moment is the only
quantity we need to know in order to fully characterize such a field. Indeed
all odd moments of a gaussian PDF are zero, while the $2n$-th moments are
proportional to the $n$-th power of the variance, $\VEV{\varepsilon^{2n}}
\propto \sigma_i^{2n}$ (the corresponding specification in Fourier space is
given in\eqref{gauss-in-k}). In the following, we shall refer to such a state
as gaussian initial conditions.

Our goal now will be to derive the properties of the PDF today, under the
assumption of a time-evolution under the sole influence of the gravitational
instability acting on gaussian initial conditions. Under the influence of
gravity, underdense regions become even more underdense (although not
indefinitely, since the density is bounded from below), while positive density
enhancements tend to grow without bound. Clearly the symmetry of the
distribution cannot be maintained, and the PDF becomes skewed, \ie
$\VEV{\delta^3}$ departs from zero. The distribution also develops a non-zero
kurtosis $\VEV{\delta^4}-3\VEV{\delta^2}^2$. In the limit of a small variance,
$\VEV{\delta^2} \ll 1$, the development of skewness and kurtosis are the most
important effects. We shall concentrate on those first.

\subsection{Moments of the PDF}

\subsubsection{Skewness of the Unsmoothed Density field:}

by using the perturbative results in the weakly-nonlinear regime, we can
calculate the gravitationally induced skewness under the assumption that
$\varepsilon$ is a random gaussian field. The lowest order terms in the series
for $\VEV{\delta^3}$ are
\begin{equation}
        \VEV{\delta^3} = \VEV{\delta^{(1)\,3}} + \VEV{ 3\, \delta^{(1)\,2}
        \delta^{(2)} } + O(\varepsilon^5) \ .
\label{delta3}
\end{equation}
The linear solution implies that the first term is $D(t)^3$ times the initial
skewness, which is zero for gaussian initial conditions. Similarly, at linear
order, all higher order moments will be proportional the initial ones; thus a
normal PDF remains normal in linear theory. The second order
solution\eqref{e2_solution} shows that the second term involves
$\VEV{\varepsilon^4}$, which is $\propto \sigma^4$ for gaussian initial
conditions. Thus the skewness ratio
\begin{equation}
        S_3 \equiv { \VEV{\delta^3} \over \VEV{\delta^2}^2 }
\label{def_s3}
\end{equation}
is a constant versus scale, and one finds (ref.~\cite{omega}) up to terms of
order $\varepsilon^2$ (since $\VEV{\varepsilon^5}=0$),
\begin{equation}
        S_3 = 4 + 4 \kappa(\Omega) \simeq {34 \over 7} +
        {6\over7}(\Omega^{-2/63} -1)\ .
\label{s3_D}
\end{equation}
The $\simeq$ sign above applies to the range of applicability of
equation\eqref{kappa-omega}.  The first term of this equation, 34/7, had been
obtained by Peebles more than a decade ago (LSS, \S18). The weak
$\Omega$-dependence of the full expression shows that nearly all the
$\Omega$-dependence of the skewness $\VEV{\delta^3}$ comes from that of the
square of the variance. It simply reflects the fact that the second order
growth rate is nearly equal to $D(t)^2$, as can be seen
from\eqref{e2_solution}.

\subsubsection{Smoothing the density field:}

to make contact with observables, we want the skewness of the density field
$\delta_\ell$, when {\it smoothed\/} with either a top-hat or a gaussian
(spherically symmetric) window, which satisfies
\[
        \int W(x)\,d^3x = 1, \qquad{\rm and}\quad \int W(x)\,x^2\,d^3x =
        \ell^2\ .
\]
The first equation insures a proper normalization to unity, while the second
requires the effective half-width $\ell$ to be finite. The top hat case is
appropriate for comparisons with the observed frequency distribution of
galaxies. (In that case, discreteness corrections must be taken into account
before comparing with the theory, \eg $\VEV{(N/\bar{N}-1)^2}=1/\bar{N} +
\VEV{\delta^2}$ if one adopts the Poisson model, see \eg LSS \S33. A gaussian
window, on the other hand, removes these discreteness fluctuations, since it
does not have sharp edges.) Here we recall the main results of~\cite{window}.

The calculations are most conveniently done in Fourier space. For the initial
field $\varepsilon$, the Fourier components are given by
\[
        \varepsilon_{\bf k} \equiv {1 \over (2\pi)^{3/2} } \int \varepsilon
        ({\bf x}) \exp (i{\bf k \cdot x}) \,d^{3}x\ .
\]
For gaussian initial conditions, one has
\begin{equation}
\begin{array}{l}
        \VEV{ \varepsilon_{\bf k} \varepsilon_{\bf k'} } = \delta_{D}({\bf k +
        k'}) P(k) , \quad\quad \VEV{ \varepsilon_\bk \varepsilon_{\bf k'}
        \varepsilon_{\bf q} } = 0 , \\ \\ \VEV{\varepsilon_{\bf k}
        \varepsilon_{\bf k'} \varepsilon_{\bf q}\varepsilon_{\bf q'} } =
        P(k)P(q)\delta_D({\bf q + q'})\delta_D({\bf k + k'}) + {\rm
        cycl.\;(2\,terms)}\ ,
\label{gauss-in-k}
\end{array}
\end{equation}
where $\delta_{D}$ is the Dirac delta, and $P(k)$ is the (initial) power
spectrum. Now we can use\eqref{delta3} and\eqref{e2_solution} to derive $S_3$
to lowest non-vanishing order,
\begin{equation}
        S_3 = \int{d^3k\,d^3k'\over(2\pi)^6\,\sigma^4} \;P(k) \, P(k')\, W_k\,
        W_{k'} \, W_{|{\bf k-k'}|} \, T({\bf k,k'}) + O(\sigma^2) \; .
\label{s3_kint}
\end{equation}
Here $T({\bf k,k'})$ stands for $T({\bf k,k'}) = 4 + 4\kappa(\Omega) - 6 \mu
(k/k') + \left[ 2 - 4\kappa(\Omega)\right] \, P_2(\mu)$, where $\mu = {\bf k
\cdot k'}/kk'$, and $P_2$ is a Legendre polynomial.  If there is no smoothing
($W_\bk = 1$), the dipole and quadrupole terms integrate to zero, and one
simply recovers\eqref{s3_D}. On the other hand, as soon as one introduces
smoothing, the result {\em does depend on the initial power spectrum} $P(k)$.

Let us assume that $P(k)\propto k^n$. Then, for a top-hat smoothing, and $-3
\le n < 1$, the equation\eqref{s3_kint} yields after painful calculations the
simple result
\begin{equation}
        S_3 = 4 + 4\kappa(\Omega) - (3+n) \; .
\label{s3_tophat}
\end{equation}

Actually, a careful inspection of the expression\eqref{s3_kint} shows that
$S_3$ should only depends on the effective (logarithmic) slope of the power
spectrum at the smoothing scale. This was confirmed in \cite{window} (see also
figure~\ref{fig-s3-cdm}) by comparing numerical integration for a CDM power
spectrum with a prediction using\eqref{s3_tophat}. The result above was
generalized in \cite{smoothpdf} to an arbitrary gaussian field, smoothed with a
top hat filter on a scale $\ell$,
\[
        S_3(\ell) = 4 + 4\kappa(\Omega) - \gamma_\ell, \quad {\rm with\quad }
        \gamma_\ell = - \partder{ \log\VEV{\delta^2} }{\log \ell} .
\]
For a pure power law $P(k)$, we have $\gamma = 3 +n$, in agreement
with\eqref{s3_tophat}.

The rather messy calculations involved in computing the case of smoothing by a
gaussian were recently completed in~\cite{kurt_lok} who express it in terms of
hypergeometric functions:
\begin{equation}
        S_3 = 3 \ _2F_1 \left( m, m, \frac{3}{2}, \frac{1}{4} \right) - \left(n
        + \frac{8}{7} \right)\ _2F_1 \left( m, m, \frac{5}{2}, \frac{1}{4}
        \right)
\label{s3_guauss}
\end{equation}
with $m=\frac{n+3}{2}$. The behavior of this expression with the spectral
indices $n$ and $\Omega$ is in fact very similar to that of the top-hat case,
see figure~\ref{fig-s3-z} below.

\subsubsection{Skewness of the Velocity Field PDF}

The continuity equation gives the divergence of the velocity field in terms of
the time derivative of the density contrast. Let us call $T_3$ the skewness
ratio for the divergence of the velocity field in units of the Hubble constant
($\theta = \nabla \cdot \bv /H$). One then finds~\cite{t3} for a top-hat
smoothing,
\begin{equation}
        T_3 \equiv{ \VEV{\theta^3} \over \VEV{\theta^2}^2 } = - \Omega^{-0.6}
        \left[ {\textstyle 26\over7} - \gamma_\ell \right] .
\label{t3}
\end{equation}
The corresponding (complicated) expression for the a gaussian window has
recently been obtained (together with the corresponding $S_3$) in
\cite{kurt_lok}.

The asymmetry in the distribution of $\theta$ is directly related to the
asymmetry in $\delta$: voids and clusters in the mass distribution correspond
to sources ($\theta > 0$) and sinks ($\theta < 0$) in the velocity field. This
quantity has the same dependence in $\Omega$ than the ratio of the predicted
versus observed velocity in density-velocity field comparisons, but it uses an
intrinsic property of the field, thereby bypassing reference to the density
field and to an unknown biasing parameter.

It turns out that $\theta$ may be advantageously measured by using Voronoi or
Delaunay tesselations of space~\cite{voronoi}. It remains to be seen whether
these methods provide accurate and unbiased estimators of $T_3$ when the number
of sampling points is comparable to those of observed catalogs, and one has
only the line-of-sight velocity (\ie bypassing the use of a POTENT-like
machinery). First analyses~\cite{th-hivon} are encouraging in that
regard. Finally, let us note that an analytical expression for $T_3$ in the
gaussian smoothing case has recently been obtained (together with the
corresponding $S_3$) in \cite{kurt_lok}.

\subsubsection{Kurtosis}

Using similar techniques to those described above, one can compute the Kurtosis
coefficient
\[
        S_4 = \frac{ \VEV{ \delta^4} - 3 \sigma^4 }{\sigma^6}
\]
and the corresponding $T_4$ for the expansion scalar $\theta$. One
finds~\cite{smoothpdf} for a top hat smoothing
\[
        S_4 = \frac{60712}{1323} -\frac{62}{3} \gamma_\ell + \frac{7}{3}
        \gamma_\ell^2, \quad T_4 = \frac{1}{f(\Omega)} \left[ \frac{12088}{441}
        -\frac{338}{21} \gamma_\ell + \frac{7}{3} \gamma_\ell^2 \right] .
\]
This was actually derived an equation for the evolution of the full generating
function (see below). The corresponding expressions for a Gaussian smoothing
can be found in~\cite{kurt_lok}.

\subsection{Edgeworth Expansion of the PDF}

As was shown in \cite{edgeworth}, the previous results on the moments of the
PDF in the weakly non-linear regime can be used to examine how gravitational
instability drives a PDF away from its initial state. Let us start with the
density field PDF, $P(\delta, \ell)$, or rather $p(\nu)$, the PDF of the
density field in terms of the standardized random variable $\nu \equiv
\delta_\ell / \sigma_\ell$. Since we want to describe an evolution from
gaussian initial conditions, it makes sense to consider an expansion of
$p(\nu)$ in terms of $\phi(\nu) = (2\pi)^{-1/2}\exp (-\nu^2/2)$ and its
derivatives. The {\it Gram-Charlier series} (Cram{\'e}r 1946 and references
therein) provides such an expansion:
\begin{equation}
        p(\nu) = c_0\,\phi (\nu) + {c_1\over 1!}\,\phi^{(1)}(\nu) + {c_2\over
        2!}\,\phi^{(2)}(\nu) + \ldots \; ,
\label{p-nu-n}
\end{equation} 
where $c_m$ are constant coefficients. Superscripts denote derivatives with
respect to $\nu$:
\begin{equation}
        \phi^{(m)}(\nu) \equiv {d^{m}\phi \over d\nu^{m}} = (-1)^{m}\,
        H_{m}(\nu)\, \phi(\nu),
\label{phi-nu}
\end{equation} 
where $H_{m}$ is the Hermite polynomial of degree $m$.  The Hermite polynomials
satisfy orthogonality relations (e.g. Abramowitz \& Stegun 1964), $
\int_{-\infty}^{\infty} \; H_{m}(\nu)\;H_p(\nu)\;\phi(\nu)\; d\nu = 0$, if $m
\ne p$, and $ = m\,!$ otherwise. Therefore, multiplying both sides of
equation\eqref{p-nu-n} by $H_m$ and integrating term by term yields
\begin{equation} 
        c_{m} = (-1)^{m}\;\int_{-\infty}^{\infty}\; H_{m}(\nu)\; p(\nu)\;d\nu
        \; .
\label{def-c-m}
\end{equation} 
Equation\eqref{def-c-m} gives $c_0 = 1, \; c_1 = c_2 = 0$, while for the next
four coefficients in the series we obtain
\begin{equation} 
        c_{m} = (-1)^{m}\,S_{m}\sigma_\ell^{m-2} \, , \; \; {\rm for}\; 3\leq m
        \leq 5\; ; \quad c_6 = S_6 \sigma_\ell^4 + 10 S_3^2 \sigma_\ell^2 \;.
\label{c-m}
\end{equation}
Thus the $S_m(\ell)$ have both a dynamic and a static application: they
describe the time evolution of moments of the PDF at a fixed smoothing scale
$\ell$, and they also describe the relation between moments of the PDF at a
fixed time on different smoothing scales.  In the latter case, one must also
include the scale-dependence of the $S_{m}$ if the initial power spectrum is
not scale-free.

We have seen that perturbation theory and gaussian initial conditions imply
that $\VEV{\delta_\ell^3} \propto \sigma_\ell^4$, and $S_3$ is therefore an
``order unity'' quantity when $\sigma_\ell$ is the ``small'' parameter. The
same is true for all remaining reduced moments, $S_m = O(1)$ for all $m$
\cite{bernardeau92} \& \cite{fry84}. It implies that the Gram-Charlier series
{\it is not} a proper asymptotic expansion for $p(\nu)$. In an asymptotic
expansion, the remainder term should be of higher order than the last term
retained.  However, if we truncated the series\eqref{p-nu-n} at the
$\phi^{(4)}$ term, which is $O(\sigma^2)$, we would miss another $O(\sigma^2)$
contribution coming from $c_6$ (equation\eqref{c-m}). In order to deal with
this problem, let us rearrange the Gram-Charlier expansion by collecting all
terms with the same powers of $\sigma$. The result is the so-called {\it
Edgeworth series}, with the first terms given by
\begin{equation} \hskip -5pt
        p(\nu) = \phi(\nu) - {\sigma \over 3!} S_3\phi^{(3)}(\nu) + \left[
        {1\over 4!}S_4\phi^{(4)}(\nu) + {10\over 6!}S_3^2\phi^{(6)}(\nu)
        \right] \sigma^2 + O(\sigma^3) \; .
\label{edgeworth-series}
\end{equation} 
Cram{\'e}r \cite{cram}) lists the Edgeworth series to higher order, and he
proves that it is a proper asymptotic expansion. This proof is directly
relevant to our purposes, since it implies that there are no additional
$O(\sigma^2)$ terms hiding in the Gram-Charlier series at $m > 6$. The
Edgeworth series thus provides a series expansion for the evolving PDF in
powers of the r.m.s. fluctuation $\sigma$.

Of course, if the variable we are interested in is $\theta$ instead of
$\delta$, one just adopts the appropriate analogues of $S_{m}$, \eg $S_3
\longrightarrow T_3 \equiv \VEV{\theta^3}\VEV{\theta^2}^{-2}$, and so
on. Similarly, if we are interested in the PDF shape in redshift space, one
just needs to use the corresponding analogue of $S_{m}$, $S_m^z$, see
\S\ref{par-redshift}). The Edgeworth series may also be used to relate $S_3$ to
other measures of asymmetry like $\VEV{\delta |\delta|}$ which, according to
\cite{nus_dek} may offer better signal to noise ratio than $\VEV{\delta^3}$
when applied to real galaxy surveys (by being less sensitive to the tail of the
PDF, \ie to rare events). This latter quantity is not easy to compute directly
by perturbation theory (see the appendix of \cite{edgeworth}), but it is
trivial to obtain
\begin{equation}
        \VEV{\delta |\delta|} = \sqrt{ {2\over 9\pi} } S_3 \sigma^3 +
        O(\sigma^5) \; .
\label{dabsd}
\end{equation} 
by using the Edgeworth series.

\subsection{Validity of Perturbation Theory} 

The previous results assume that the system never gets to be strongly
non-linear, \ie $\sigma_\ell \ll 1$ is true at {\em every} scale. But in the
real universe, we know that the observed density contrast is very large at
small scales. One only gets $\sigma_\ell \ll 1$ for $\ell \gg 8h^{-1}$ Mpc. It
is thus by no means obvious that there is any range today for which
perturbation theory might be applicable.  On the other hand, it has long ago
been noticed that linear perturbation theory yields a good description of the
variance [or the 2-body correlation function $\xi(t) \propto D(t)^2/D(t_i)\:
\xi(t_i)$] measured in fully non-linear numerical simulation.

\begin{figure}[hbtp] \hbox{
\psfig{file=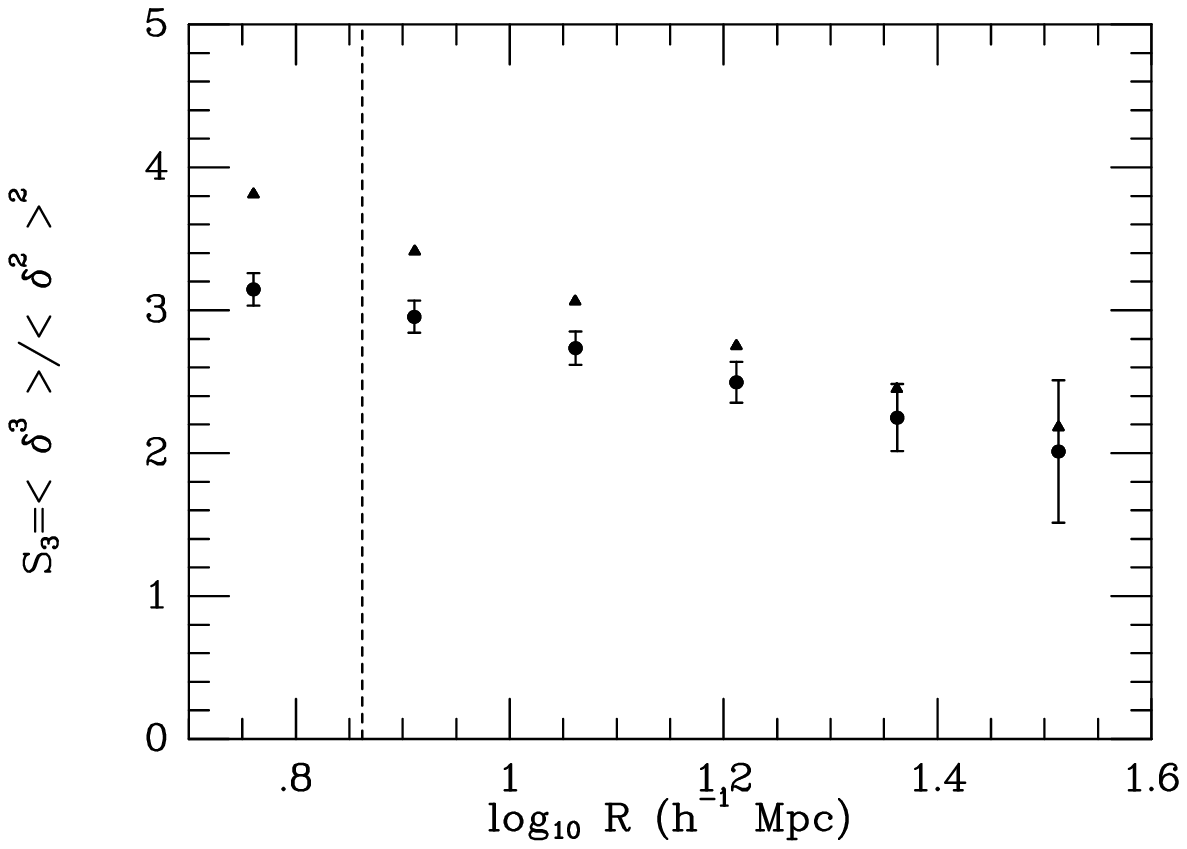,width=0.5\textwidth,bbllx=76pt,bblly=35pt,bburx=414pt,bbury=275pt}
\psfig{figure=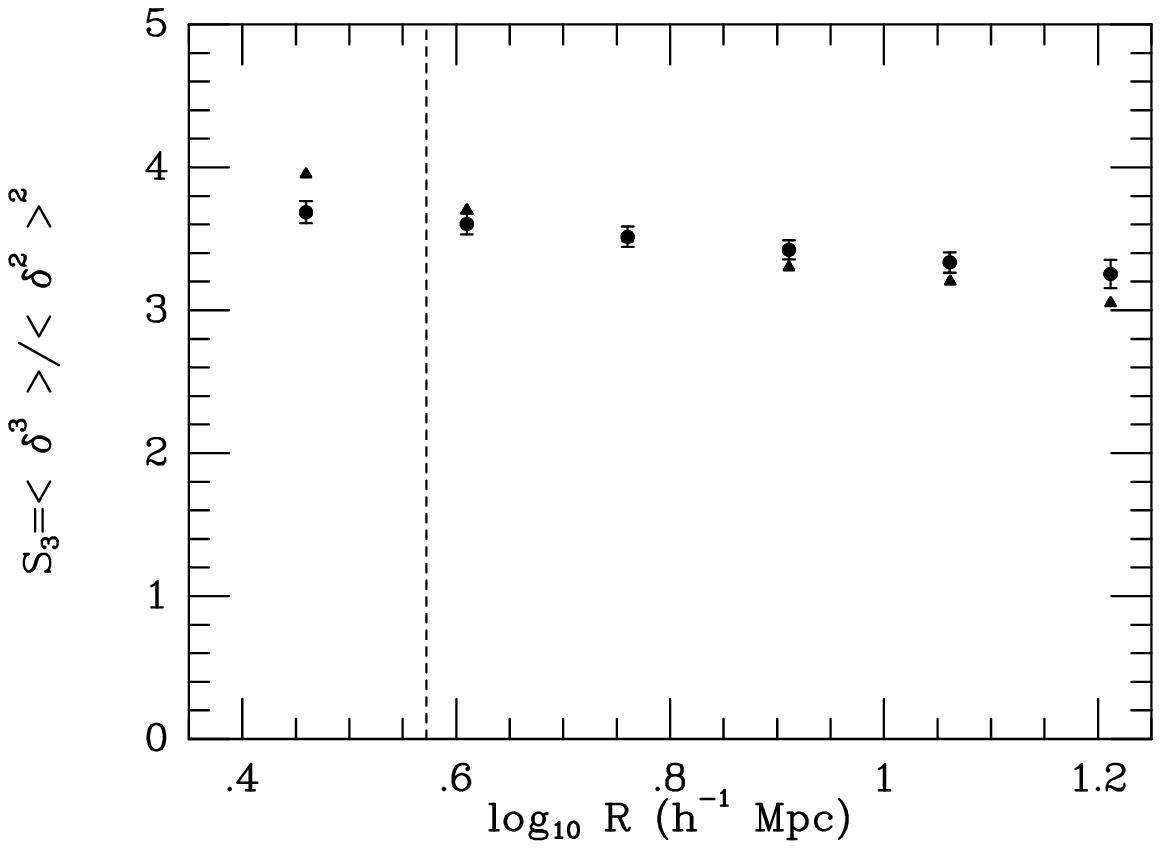,width=0.5\textwidth,bbllx=76pt,bblly=35pt,bburx=414pt,bbury=275pt}
}
\caption{The measured $S_3(\ell)$ in a CDM simulation (triangles) is compared
to the theory (circles; the error bars come form the numerical uncertainty in
evaluating the integrals giving $S_3$). The left panel corresponds to a
spherical top-hat smoothing, while the right one corresponds to a gaussian
smoothing. The vertical dashes mark the limit between the strongly non-linear
regime ($\sigma \geq 1$) and the weakly non-linear one ($\sigma \leq
1$). Courtesy Colombi 1993 \& Juszkiewicz \etal 1993a.
}
\label{fig-s3-cdm} 
\end{figure} 

It turns out that what holds true for linear perturbation theory results is
also true at higher orders. Indeed, the top-hat expressions for $S_3$ have been
checked against the simulation results in \cite{edfw}, \cite{compare} \&
\cite{lahav}, in the case of scale-free initial spectra. In the case of a
gaussian smoothing, the theory was checked with the scale free-simulations in
\cite{wein_cole}. In the CDM case, comparison were made with the results in
\cite{compare}, as well as simulations performed for that purpose, see figure
\ref{fig-s3-cdm}. In all those cases, the agreement between the perturbation
theory and N-body experiments was excellent (see \cite{window} for further
details), up to surprisingly large values of the variance $\sim 1$.

\begin{figure}[hbtp] 
\hbox{ \hskip -5pt
\psfig{file=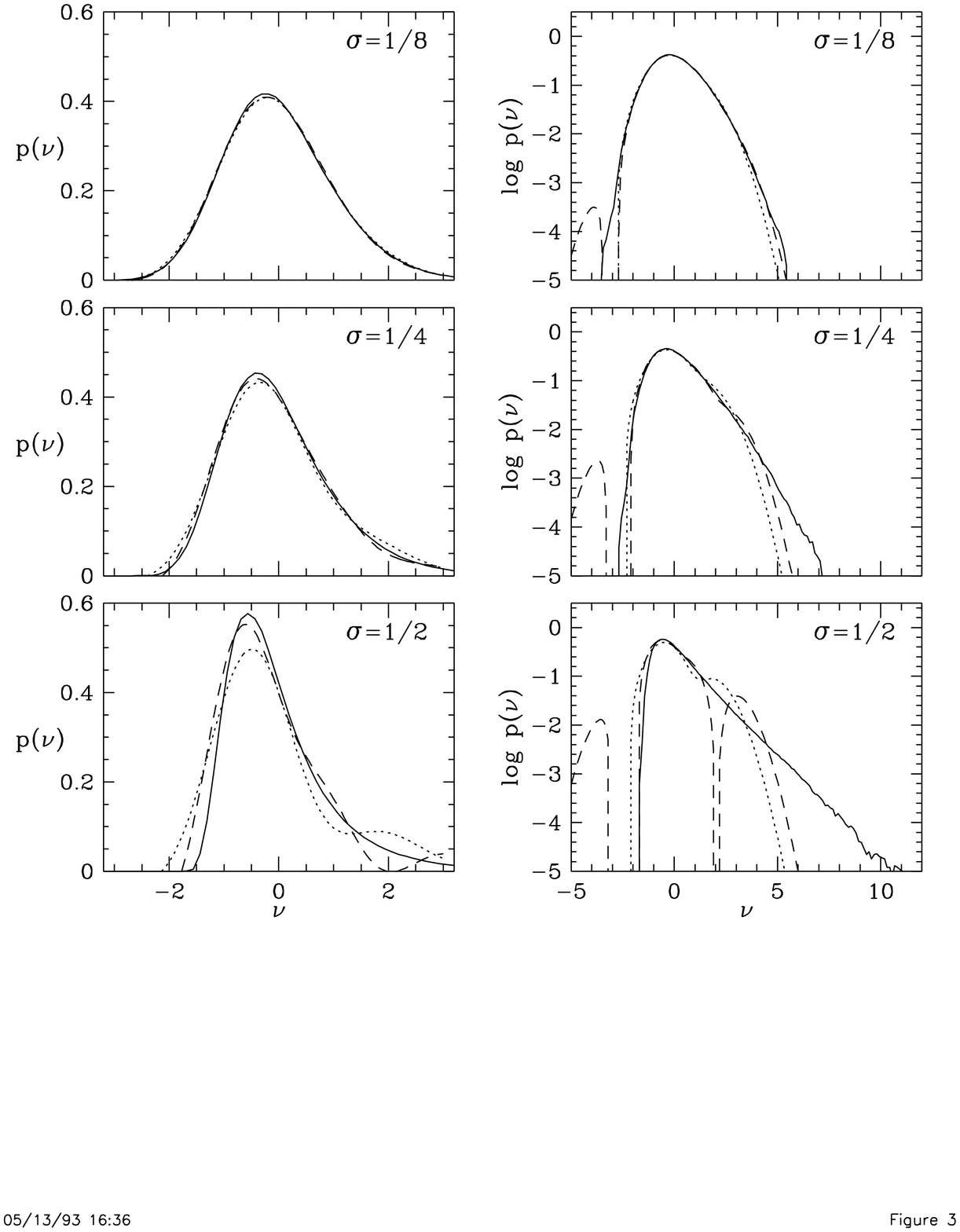,width=0.5\textwidth,clip=t,bbllx=318pt,bblly=196pt,bburx=575pt,bbury=745pt}
\psfig{file=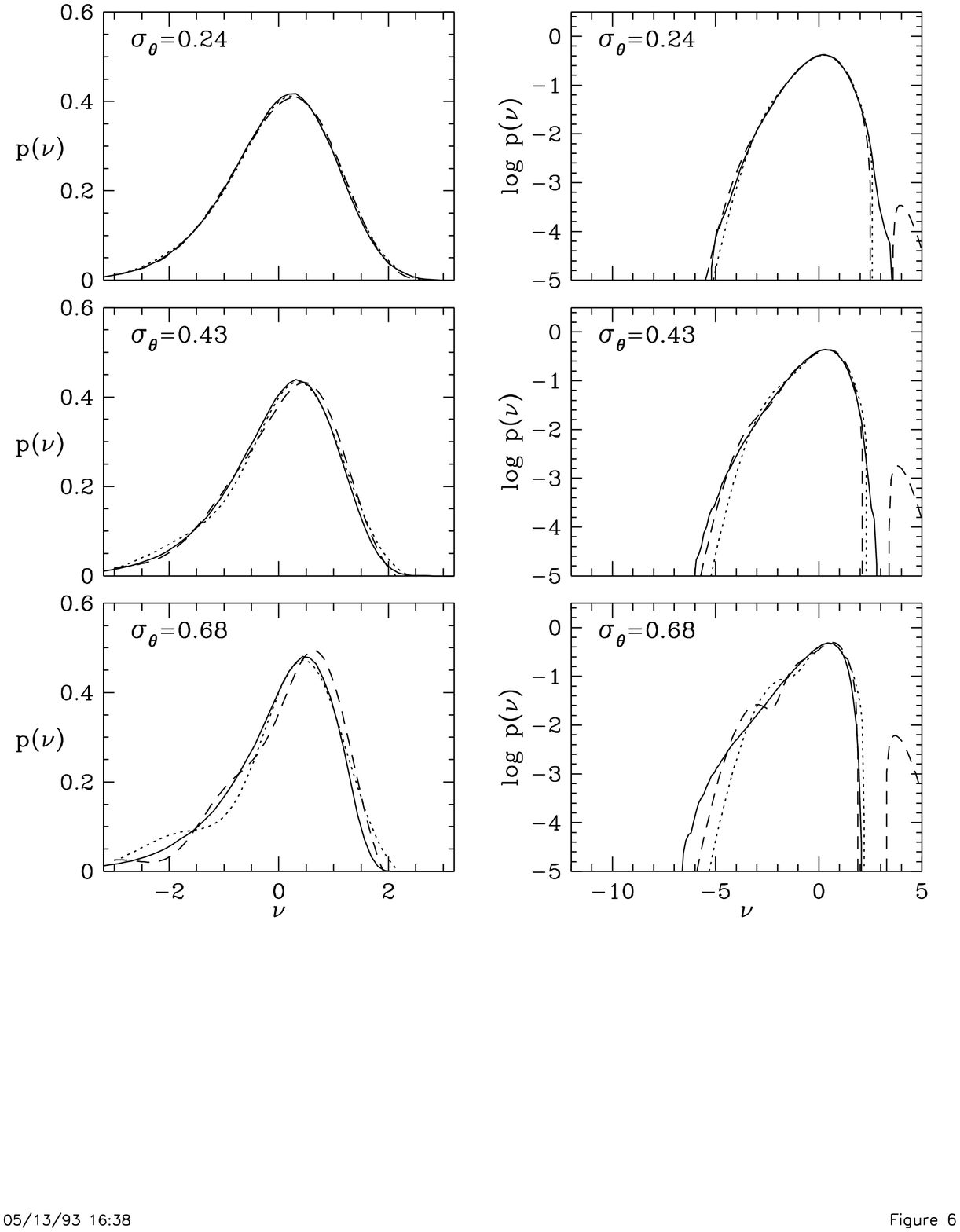,width=0.5\textwidth,clip=t,bbllx=318pt,bblly=196pt,bburx=575pt,bbury=745pt}
}
\caption{Comparisons of the Edgeworth PDF with simulations (solid lines). See
details in the text. Courtesy Juszkiewicz \etal 1993b.
}
\label{fig-p-delta-theta}
\end{figure} 

One can also check when the Edgeworth series and perturbation theory are
appropriate to describe the overall shape of the PDF. The solid lines of figure
\ref{fig-p-delta-theta} show the PDF of the smoothed density fields of matter
density (left column) and velocity divergence (right column), at three
different stages in scale-free n=-1 simulations. The corresponding {\em rms}
amplitude ($\sigma$ and $\sigma_\theta$) are shown in each panel, and the
variables on the horizontal axes are $\nu = \delta / \sigma$ and $\nu = \theta
/ \sigma_\theta$. In all the panels, dotted and dashed lines show the
approximations keeping the terms of order $\sigma$ and $\sigma^2$ respectively
in eq.\eqref{edgeworth-series}. The approximations work well for $S_3 \sigma <
1$ and $|\nu \sigma| < 1$ (and equivalent requirements for $\theta$). They
begin to break down outside of that range, as expected.

So far, all the comparisons done lead to the same conclusion: perturbation
theory predicts remarquably accurately the simulation results when $\sigma
\simlt 1$ ($|\nu \sigma| < 1$ for the Edgeworth PDF), even when small scales
are strongly non-linear. All the discrepancies found could actually be traced
to limitations of the simulations themselves. What remains to be found though,
is why perturbation works so well, even at scales where the density contrast is
comparable to unity.

\section{FURTHER DEVELOPMENTS\label{sec:further}}

\subsection{Higher Order Moments}

While appropriate to break fresh grounds, the pedestrian calculation approach
used so far would be quite unwieldy to compute higher order
moments. Fortunately, Bernardeau~\cite{bernardeau92} found an elegant shortcut
by means of a dynamical equation for the generating function of quantities
closely related to the moments. This function simply obeys the evolution
equation of the spherical collapse. One nice aspect of this approach is that it
also gives the moments hierarchy for approximate descriptions of the dynamics
(like those described in \S~\ref{sec:anl} and afterwards). Indeed, in that
case, the moments are then obtained from the generating function obeying the
evolution equation of the spherical collapse, as given by the considered
approximation~\cite{munshi}. In addition, this approach was successfully
generalized to account for a top-hat smoothing~\cite{smoothpdf}. Unfortunately,
this generalization uses a trick specific to the top-hat case. Gaussian
smoothing calculation still have to be done ``by hand''.  used for
 
\subsection{Higher Order Corrections to the Moments}

All the calculations presented so far were performed retaining only the leading
non-linear corrections, that is in the limit $\sigma \longrightarrow
0$. Recently, though, the next higher order correction to all moments was
obtained~\cite{scocci}, albeit in the absence of smoothing. These (one-loop)
corrections start to dominate the leading non-linear (tree-level) corrections
when $\sigma \sim 1/2$ for the density field and $\sigma \sim 1$ for the
divergence of the velocity field.

\begin{figure}[hbtp] \centering \centerline{ \hbox{
\psfig{file=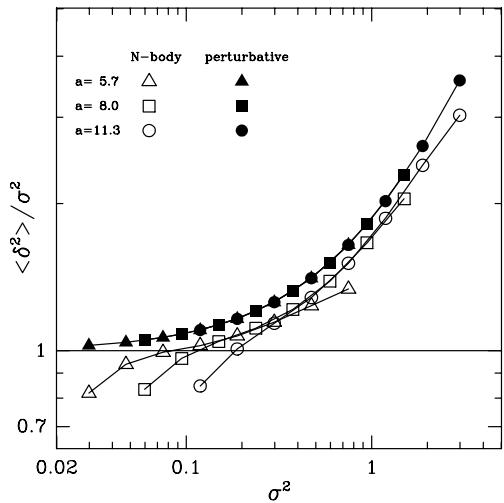,width=0.5\textwidth,bbllx=29pt,bblly=148pt,bburx=563pt,bbury=687pt}
} }
\caption{Previrialisation effect in the $n=-2$ case. Courtesy Lokas \etal\
1995b.}
\label{fig:previr}
\end{figure} 

The only case so far when these one-loop correction were obtained in the
presence of smoothing is for the simplest moment, the variance
itself~\cite{previr}. One can then approach the long-standing
``previrialisation'' problem: do the presence of small scale slows down the
collapse of larger structures encompassing them? Or in other words, do
non-linear interactions between different scales increase or decrease the rate
of growth of structures. Perturbation theory can at least give us a hint of
what the first non-linear corrections do. Lokas \etal \cite{previr} find that
the weakly non linear growth of variance is very similar to the linear one,
when the index $n$ of the primordial power spectrum is close to -1, in
agreement with the numerical simulation results of \cite{evrard_crone} who
concluded to the absence of a previrialisation effect. But it is decreased if
it is shallower, i.e.  if small scale structures are more prominent. This does
bring support to Peebles' (very educated) guess when he argued that small scale
structures will promote the development of non-radial motions in the larger
collapsing region, these motions acting effectively as a stabilizing pressure
term. Once again, those perturbative results are confirmed by simulations
results, as can be seen on figure~\ref{fig:previr}.

\subsection{General Initial Conditions}

So far, we focused our attention to the case of Gaussian initial conditions for
which the reduced moments, or cumulants, are initially zero, and thus $S_3$ and
$S_4$ too. We found that in this specific case, the $S_3$ and $S_4$ which
develop under the influence of gravity take on particular {\em constant}
values. But what happens for more generic initial conditions? Can the
predictions for the Gaussian case can be used as observational tests of the
Gaussian hypothesis?

Initially, one expects on dimensional grounds to have $\VEV{\delta^3} \propto
\sigma^3$, i.e. $S_3 \propto 1/\sigma$. But non-gaussian fluctuations will also
built a gravitationally induced contribution during their evolution, which may
quickly dominate the initial term. This was demonstrated in specific cases by
\cite{coles} \& \cite{luo}, and \cite{fry-scherrer} succeeded in performing the
calculation for arbitrary initial conditions. They found that $s =
\VEV{\delta^3}/\sigma^3$ is equal to its initial value, $s_i$ plus a term
proportional to $\sigma$, the coefficient of proportionality depending on the
assumed initial conditions. Unfortunately, biasing (see below) may change this
proportionality constant, thus comparisons to data will only put an upper limit
on $s_i$.

Fortunately, as shown in \cite{kurt_chod}, similar calculation at the next
order show that $k = \VEV{\delta^4}/\sigma^4$ is equal to its initial value
$k_i$ plus a term linear in $\sigma$. This term is zero in the gaussian case
(the first non-zero term is in $\sigma^2$, leading to a constant $S_4$). This
measure should thus be a more powerful probe of the gaussian hypothesis, since
then this is the scaling with $\sigma$ which is altered in the generic case,
not a coefficient in front. The gravitational evolution is then less likely to
cast a shadow on the signature of non-gaussian initial conditions.

\subsection{Biasing and comparisons to data}

\begin{figure}[hbtp] \centering \centerline{\hbox{
\psfig{file=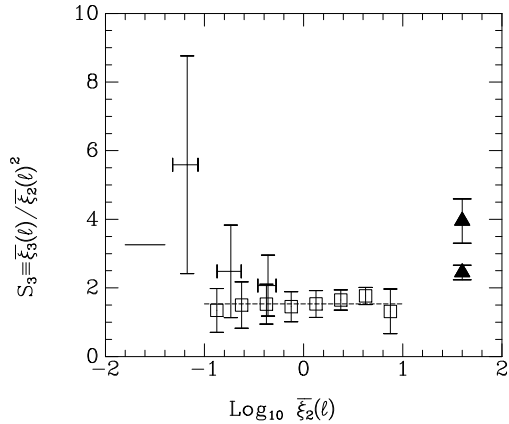,bbllx=44pt,bblly=201pt,bburx=560pt,bbury=640pt,width=0.5\textwidth}}}

\caption{Skewness measurement in the \iras\ 1.2 Jy catalog. Courtesy Bouchet
\etal 1993)} \label{fig-s3-iras}
\end{figure}

As was mentioned earlier, it might be that galaxies are not fair tracers of the
mass, even at the large scales where perturbation theory is applicable. But let
us suppose that this biasing is local, \ie that the galaxy field at scale
$\ell$, $\delta_{g, \ell} $ is some unknown function $B$ of the underlying
field at the same location, $\delta_{g, \ell}(x) = B[\delta_\ell(x)]$.  Then a
Taylor expansion would give
\[
        \delta_g = \sum_{k=0} \frac{b_k}{ k! } \delta^k ,
\] 
where $b_0$ is set so that $\VEV{\delta_g}=0$, $b_1$ is the {\em linear} bias
introduced previously, and $b_k = {\rm d}^n B/{\rm d}y^n|_{y=0}$.  Replacing
this into the expression for skewness, one sees that the galactic field
skewness, $S_3^g$, is given by~\cite{edgeworth}
\begin{equation}
        S_3^g = {S_3 \over b_1} + {3b_2 \over b_1^2} .
\label{gal_skew}
\end{equation} 
If the biasing is not scale-dependent (why should it be on very large scales?),
then the constancy with scale of $S_3$ is preserved. This agrees well with the
determination~\cite{iras} in the \iras\ 1.2Jy catalog (see
fig.~\ref{fig-s3-iras}) or in the SSRS~\cite{gaz-yok}. In addition, if a linear
bias is assumed, like it is customary in density-velocity comparisons for
measuring $\beta = \Omega^{0.6} / b_1$, then one gets the linear bias parameter
$b_1$, {\em independently} of the (poorly known!) value of $\Omega$. Indeed,
once the shape of the window function is fixed, $S_3$ depends in a known way on
the local slope of the power spectrum only,which can be determined
self-consistently by measuring the slope of the second moment in the same
catalog.

Of course, there might be a non-linear biasing term, $b_2$, and a skewness
measurement alone will then only constrain the combination\eqref{gal_skew}. One
cannot hope to break this degeneracy by measuring the galactic field kurtosis,
since then a new unknown, $b_3$, would enter the game. On the other hand, one
can compare measurements in different catalogs, e.g. optically and infrared
selected, to obtain ratios of biasing parameters \cite{fry-gaz} (see also
\cite{lah_sas}). In addition, by extending systematically considerations
similar to those above, \cite{fry-gaz} develop a general (local) bias
transformation theory and show that the signature of gaussian initial
conditions evolving under gravity, i.e. $S_n$ = constant is preserved by a
local biasing, i.e. $S_n^g$ remains a constant (with a different value) at all
orders $n$. On the contrary, as was shown in~\cite{coopgal}, a cooperative
galaxy formation scheme can be severely constrained by the skewness it would
predict.

Finally, let us mention that a similar scaling of the moments with the second
one has been found in the APM survey \cite{gaz-apm}, which is indeed
expected~\cite{bern_2d} theoretically.


\section{ZELDOVICH APPROXIMATION\label{sec:zeldo}}

As was shown above, a lot can be accomplished by using the standard Eulerian
perturbative approach to gravitational instability.  However, perturbative
equations are often easier to integrate when expressed in Lagrangian
coordinates. In our experience, this happened at second order for $\Omega \not=
1$ (see \S\ref{par-lag-form} below). Another example is the redshift space
distortion for $\VEV{\delta^3}$, discussed in \S\ref{par-redshift}. Moreover,
as $\VEV{\delta^2}$ grows with time, at any fixed order in perturbation theory,
the Lagrangian approach is likely to remain valid longer than the Eulerian
approach. This is so because the requirement of small Lagrangian displacements
and gradients is weaker than the requirement $\VEV{\delta^2} \ll 1$. This idea,
which motivated Zel'dovich \cite{zeldo} approximation (which is simply the
extrapolation of the first order Lagrangian solution in a regime where its
validity is not mathematically warranted), remains valid at higher orders.

In the following, I first introduce Zel'dovich approximation by using previous
results concerning the linear Eulerian theory. I then discuss an application to
the growth of galactic spin (tidal torque theory), as well as the idea of
``Pancakes''.

In a second step, I outline the general derivation of the perturbative
(Lagrangian) solutions, and focus on the redshift distortion problem as an
application. I also give examples of comparison between Eulerian and Lagrangian
theory when $\VEV{\delta^2} \simeq 1$, i.e when Zeldovich and higher order
approximations are used as approximations to the real dynamics, in a regime
where there validity is hard to assess analytically.

\subsection{Derivation from Eulerian Theory}

We found above that in linear Eulerian theory the evolution of the growing mode
is self-similar, with
\[
        \delta(\bx, \tau) = D(\tau)\varepsilon(\bx), \quad \phi(\bx, \tau) = {D
        \over a} \phi_i(\bx); \quad \nabla^2\phi_i = 4 \pi G \rho_b a^3
        \varepsilon(\bx) ,
\] 
the subscript $i$ referring to an ``initial'' state. Note that in an
Einstein--de Sitter Universe, $D(\tau) \propto a$, and thus the potential is
(linearly) conserved. In addition, the linearized Euler equation using the
conformal time $\tau$, $a\dot\bv + \dot a\bv = -a \nabla\phi = -D\nabla\phi_i$
(with $\bv = \dot\bx$) yields $\bv = -(1/a) \int D {\rm d}\tau \nabla\phi_i =
-(1/D) \int D {\rm d}\tau \nabla \phi$ (i.e. $\bv$ is parallel to
$\nabla\phi$).

We can thus use these expressions to get the linear trajectory of a particle as
a displacement field, $\bPsi (t, \bq) = \bx - \bq$, applied to its initial
position, or Lagrangian coordinate, $\bq$,
\begin{equation}
        \bx = \bq + \bPsi (\tau, \bq) \quad{\rm with}\quad \bPsi = \left[ \int
        {d\tau \over a}\int D d\tau \right] \nabla\phi_i .
\label{define_lagrange_zel}
\end{equation} 
By using the continuity equation written as $a\ddot D = \dot a D= 4 \pi G
\rho_b a^3 D$, we obtain
\begin{equation}
         \bPsi(\tau, \bq) = b(\tau) \bPsi_i, \quad{\rm with}\quad b(\tau) =
         {D(\tau)\over 4 \pi G \rho_b a^3}, \bPsi_i = \nabla\phi_i ,
\label{define_b}
\end{equation} 
the velocity field being given by $\bv = -a \dot b \bPsi_i$. At the linear
order then, particles just go straight (in comoving coordinates) in the
direction set by their initial velocity.

Zeldovich' approximation is to use these linear solutions to extrapolate fluid
element trajectories well beyond the validity range of the linear Eulerian
theory ($\delta \ll 1$).

\subsection{Galactic Spin Origin}

Zeldovich approximation allows to see quite easily how the tidal field of an
inhomogeneous distribution couples to the quadrupole of the mass distribution
of a proto-object to induce a growth of its angular momentum $\bM$. At early
times,
\[
        \bM = \int_{V} d^3q\, \rho a^3 (a\bx - a\bar{\bx}) \wedge \bv/a ,
\] 
where the integral is taken over the initial (lagrangian) volume of the
perturbation $V$, of center of mass $\bar{\bx} = \int_V d^3q \bx/V$. At the
lowest order, $\rho \equiv \rho_b$ and $\bv = -a \dot b \nabla\phi_i$, and thus
\[
        \bM = -\rho_b a^4 b \int_V d^3q\, (\bq - \bar{\bq}) \wedge \nabla
        \phi_i = - \rho_b a^4 \dot b \int_S \phi_i (\bq - \bar{\bq}) \wedge dS
        .
\] 
The second expression follows by integration par part (Gauss' divergence
theorem), and the integral is thus taken over the boundary surface of the
lagrangian volume of the future object. If this volume is spherical (or its
surface $S$ is an equipotential), then $\bM$ remains zero. Provided the initial
potential is smooth enough, we can take advantage of a Taylor expansion of the
acceleration around the center of mass,
\[
        \nabla \phi_i|_{\bq} = \nabla \phi_i|_{\bar{\bq}} + (\bq - \bar{\bq})
        \cdot \nabla\nabla\phi_i|_{\bar{\bq}} + {\cal O} \left[ (\bq -
        \bar{\bq} )^2 \right] ,
\] 
to relate the angular momentum to the inertia tensor of the proto-object,
$\cI$, and the (initial) tide tensor at the center of mass position, $\cT$
defined in\eqref{def:tide_tens},
\[
        M_i = -a \dot b \epsilon_{ijk} T_{jl} I_{lk}, \quad{\rm with}\quad
        \cI_{lk} = \int_V (q_l - \bar{q_l}) (q_k - \bar{q_k}) \rho_b a^3 d^3q ,
\] 
$\epsilon$ standing for the totally antisymmetric tensor of order 3.

In the Einstein-de Sitter case, the linear growth rate of the angular momentum
is thus
\[
        \bM \propto \tau^3 \propto t .
\] 
Of course, this description holds only during the early collapse phase ($\sim $
while the proto-object is still expanding). Later one, particle turn over, and
there is no chance that their trajectories may be described by a simple
ballistic approximation. Latter phases have been mostly studied by numerical
simulations. Still, the simple exercise above shows how Zeldovich approximation
clearly unveils the basic principle of the tidal torque theory of the galactic
spin origin, i.e. that the tidal field of neighboring density perturbation acts
on the irregular shape of the proto-object.

\subsection{Density Contrast \& Pancakes\label{sec:pancakes}}

The density contrast simply obtains from the trajectories by requiring that
mass be conserved, i.e.
\[
        \rho(\tau) d^3x = \rho(\tau_i) d^3q \Longrightarrow 1+\delta = {1\over
        J} ,
\] 
$J$ standing for the Jacobian of the transformation from $\bq$ to $\bx$:
\[
        J = \left| {\partial \bx \over \partial \bq} \right| = | det\, \cD | .
\] 
The deformation tensor, $\cD$ may be written as the identity matrix plus the
tidal tensor $\cT = |\partial \bPsi / \partial \bq |$, which is directly given
by the curvature of the initial potential field (in Zeldovich approximation, $T
= b \nabla\nabla\phi_i / (4 \pi G \rho_b a^3)$).

Let us call $\lambda_1 \geq \lambda_2 \geq \lambda_3$ the local eigenvalues of
the (initial) tidal field. The density contrast may then be written as
\[
        1+\delta = {1 \over \left[ 1 -b(\tau) \lambda_1 \right] \left[ 1
        -b(\tau) \lambda_2 \right] \left[ 1 -b(\tau) \lambda_3 \right] } ,
\] 
which shows that, if this approximation remains approximately valid till the
first crossing of trajectories, one expect a singularity, a (locally) planar
collapse to infinite density along the axis of the largest positive (if any)
eigenvalue, $\lambda_1$, when $b(\tau) \lambda_1 \rightarrow 1$. If no
eigenvalue is positive, this is a developing underdense region, while in the
rarer case when two eigenvalues are approximately equal, $\lambda_1 = \lambda_2
\not= \lambda_3$, there is a collapse to a filament. The case $\lambda_1 =
\lambda_2 = \lambda_3$ leads to a spherical collapse.

For a gaussian field, Doroshkevich found the probability distribution of the
eigenvalues, and in particular that $P(\lambda_1 > 0, \lambda_2 < 0, \lambda_3
<0) = 0.42$ ( $ = P( \lambda_1 < 0, \lambda_2 > 0, \lambda_3 > 0$ ). It shows
that indeed one generically expects ``pancakes'' around local maxima of
$\lambda_1$. This is to be contrasted with the {\em linear} Eulerian theory
whose extrapolation would lead to predicting a collapse around the initial
maxima of the density field (since $\delta \propto \delta_i$), i.e. around
local maxima of $\lambda_1 + \lambda_2 + \lambda_3$ whose collapse is rather
spherical.

The ``pancake'' theory arose in the context of a ``neutrino'' dominated dark
matter, where there is a natural small scale cut-off in the power spectrum of
perturbations due to their early phase of free-streaming. Then the potential
field is quite smooth and it is not too surprising that Zeldovich approximation
provides a quite accurate description of the large scale structures which
appear in numerical simulations. Indeed, we only have to describe an
essentially laminar flow. On the other hand, in a hierarchical scenario where
small scale structures collapse first, these small scale potential wells will
undoubtedly curve the particles trajectories and Zeldovich approximation would
not describe at all the individual particle trajectories. We shall see later
that this approximation might still be useful, but in an average sense.

\section{LAGRANGIAN PERTURBATION THEORY\label{sec:lpt}}

In the previous section, we made contact with a lagrangian description of the
dynamics by introducing Zeldovich approximation as a consequence of the
Eulerian perturbation theory at the linear order. In the following, we rather
describe the steps involved in a systematic perturbative approach of Lagrangian
theory, as initiated by Moutarde \etal 1991, \cite{moutarde} (see also the
contribution of T. Buchert in this volume). We then proceed with an application
to these results to the computing of the skewness in redshift space. We shall
later come back to using those rigorous perturbative solution as approximations
to the full non-linear dynamics, thereby extending Zeldovich ansatz.
 
\subsection{Perturbative equations\label{par-lag-form}}

The Lagrangian perturbative approach introduced by Moutarde \etal (1991)
proceeds in quite a parallel fashion to the Eulerian one reviewed earlier. As
before, we use the Newtonian approximation with comoving coordinates (see the
contribution of S.  Matarese in this volume for the full general relativistic
approach). Following Doroshkevich \etal 1973 \cite{doro}, we replace the
standard cosmological time by a new time $\tau$ defined by
\begin{equation} 
        d\tau \propto a^{-2} dt .
\label{time_transfo}
\end{equation}
Beware that we kept the same symbol here than above for the conformal time
(which is defined instead by $a d\tau \propto dt$!).  In that case the motion
and field equations now read
\begin{equation} 
        \ddot \bx = -\nabla_{x} \Phi,\quad \Delta_{x} \Phi = \beta(\tau) \delta
        ,
\label{basic_eqs}
\end{equation}
with no $\dot\bx$ term. In this equation, dots denote derivatives with respect
to $\tau$, and $\beta(\tau) = 4\pi G a (\bar\rho a^3)$. By choosing properly
the proportionality constant in the definition\eqref{time_transfo} when,
respectively, $\Omega$ is equal, smaller or greater than 1, $\beta$ is simply
given by
\[ 
        \beta = {6\over \tau^2 + k(\Omega)} ,
\] 
with $k(\Omega = 1) = 0$, $k(\Omega < 1) = -1$, and $k(\Omega > 1) = +1$. While
$\tau = t^{-1/3}$, when $\Omega =1$, one has $\tau = |1-\Omega|^{-1/2}$
otherwise.

Now we wish to follow the particle trajectories instead of the density
contrast, by using the mapping
\begin{equation} 
        \bx(\tau) = \bq + \bPsi(\tau, \bq)
\label{define_lagrange}
\end{equation}
The Jacobian of the transformation from $\bx$ to $\bq$, permits as before to
express the requirement of mass conservation simply as
\[
        \rho(\bx )\, J\, d^3q = \rho(\bq ) d^3 q ,
\]
\ie $\delta = J^{-1} -1$ (note that $\rho(\bx)$ is in fact a shorthand for
$\rho(\bq, t)$). By taking the divergence of the equation of
motion\eqref{basic_eqs}, one obtains the equivalent of the Eulerian
equation\eqref{eq:dens}:
\begin{equation} 
        J(\tau, \bq) \nabla_{x} \ddot \bx = \beta (\tau ) \left[J(\tau, \bq) -
        1 \right] .
\label{new_master}
\end{equation}
Of course the addition of any divergence-free displacement field to a solution
of the previous equation will also be a solution. In the following, we remove
this indeterminacy by restricting our attention to potential movements, which
must satisfy
\begin{equation} 
        \nabla_{x} \times \dot \bx = 0 .
\label{potential condition}
\end{equation}
The main reason to restrict to that case is that vortical perturbations
linearly decay, a consequence of the conservation of angular momentum in an
expanding universe. Thus one might consider that the solutions will apply
anyway, even if vorticity is initially present, because at later times it will
decay away. In the same spirit, we shall mainly focus on growing mode solution
(see \cite{b_ehlers, buc94} for the cases of rotational perturbations in
Lagrangian space but not in Eulerian space, and also for the effect of decaying
modes).
 
The final equation to solve obtains by rewriting the divergence of the
acceleration $\Gamma \equiv \ddot \bx$ explicitly as a function of $\bq$
\begin{equation}
        \nabla_{x} \Gamma = J(\tau, \bq)^{-1} \sum_{i,j} \Gamma_{i,j} A_{ji} ,
\label{diff_transfo}
\end{equation}
where the $A_{ij}$ are the cofactors of the Jacobian, and the partial
derivatives denoted by Latin letter are taken with respect to the $\bq$
coordinate (\eg, $\Gamma_{i,j} = {\partial \Gamma_{i}\over \partial q_j}$).

As in the Eulerian case, perturbative solutions are obtained by means of an
iterative procedure. But this time, the expansion concerns the particles
displacement field itself, and we write it as
\begin{equation} 
        \bPsi = \varepsilon\bdis{1} + \eps{2} \bdis{2} + \eps{3} \bdis{3} +
        \ord{4} .
\label{def_exp_psi}
\end{equation}
The determinant of the Jacobian is then similarly expanded as
\begin{eqnarray}
        J & = 1& + \varepsilon \jac{1} + \eps{2} \jac{2} + \eps{3} \jac{3} +
        \ord{4} \nonumber \\ & = 1& + \varepsilon \K{1} + \eps{2} (\K{2} +
        \L{2})\\ & & + \eps{3} (\K{3} + \L{3} + \M{3}) + \ord{4} \nonumber
\label{def_exp_jac}
\end{eqnarray}
where $\K{m}$, $\L{m}$, and $\M{m}$ denote the $m-th$ order part of the
(invariant) scalars
\[ 
\left\{
\begin{array}{ll} 
        K & = \nabla \cdot \bPsi = \sum_{i} \Psi_{i,i} \\ L & = {1\over 2}
        \sum_{i\not= j} ( \Psi_{i,i} \Psi_{j,j} - \Psi_{i,j} \Psi_{j,i} ) \\ M
        & = {\cal D} = \det [\Psi_{i,j}] \ .
\end{array}	
\right.
\]
In other words,
\begin{equation} 
\left\{
\begin{array}{ll} 
        \K{m} & = \nabla \cdot \bdis{m} = \sum_{i} \dis{m}_{i,i} \\ \L{2} & =
        {1\over 2} \sum_{i\not= j} ( \dis{1}_{i,i} \dis{1}_{j,j} -
        \dis{1}_{i,j}\dis{1}_{j,i} ) \\ \L{3} & = \sum_{i\not= j} (
        \dis{2}_{i,i} \dis{1}_{j,j} - \dis{2}_{i,j}\dis{1}_{j,i} ) \\ \M{3} & =
        \det [\dis{1}_{i,j}] .
\end{array}	
\right.
\label{defs}
\end{equation}
The equation to solve then follows by replacing the expansions
(\ref{def_exp_psi}-\ref{def_exp_jac}) of the displacement field and of the
Jacobian in equations\eqref{new_master} and \eqref{diff_transfo}. We now
proceed to an iterative solution of this equation.

\subsection{Low order Solutions}

I give below only the solutions for the first two orders, when $\Lambda=0$ and
$\Omega$ is either unity or smaller than one. This is all we need for the
application described afterwards. A more complete set in solution can be found
in Bouchet \etal 1995 \cite{lagrange} who give for the first three orders the
solutions for arbitrary values of $\Omega$ when there is no cosmological
constant, as well as numerical solutions for flat universes when $\Lambda \not=
0$.

\subsubsection{First Order:}

keeping only the terms of order $\varepsilon$, one gets
\begin{equation} 
        \ddot \K{1} - \beta \K{1} =0 .
\label{eq_order1}
\end{equation}
Thus $\K{1}(\tau, \bq)$ can be factorized into a temporal and spatial part,
$\K{1}(\tau, \bq) = g_1(\tau) \K{1}(\tau_i, \bq)$, where $\K{1}(\tau_i, \bq)$
is the divergence of the initial displacement field, $g_1(\tau_i)$ is assumed
to be unity, and $g_1 = k_a g_{1a} + k_b\, g_{1b}$. For $\Omega =1$, the linear
growth rate $g_1$ is given by $g_{1a} = \tau^{-2}$, and $g_{1b} = \tau^3$. When
$\Omega < 1$,
\[
        g_{1a} = 1 + 3 (\tau^2 - 1) (1+ \tau S),\quad g_{1b} = \tau (\tau^2 -1)
        ,\quad S = {1\over 2}\ln {\tau -1 \over \tau +1} \ .
\]
The closed case is given in \cite{lagrange}.

For a potential movement, one thus recovers Zeldovich solution, $\bdis{1}
(\tau, \bq) = g_1(\tau )\bdist{1} (\bq)$, as in \eqref{define_lagrange_zel}, if
we define (for all $m$)
\[ 
        \bdist{m}(\bq)\equiv \bdis{m}(\tau_i, \bq) .
\] 
Also, note that $\del{1}(\tau ) \propto g_1(\tau )$ (with initially $\del{1}_i
= - \nabla \cdot \bdist{1} (\bq)$), \ie, the Eulerian linear behavior is
recovered, since $g_1 \equiv D$.

The logarithmic derivative of the growth factor, $ f_1 \equiv \frac{a}{g_1}
\frac{{\rm d} g_1}{{\rm d}a}$, is useful to describe comoving peculiar
velocities and therefore redshift distortions (see below). Near $\Omega=1$, a
limited expansion of the solution shows that $f_1\asymp \Omega^{4/7}$ where
$\asymp$ means ``behaves asymptotically as''. A better analytical fit for $f_1$
in the range $0.1 < \Omega < 1$ is given by $ f_1\approx \Omega^{3/5}$, as was
originally proposed in \cite{P76}.

\subsubsection{Second Order:}

collecting terms of $\ord{2}$ in equations\eqref{new_master} and
\eqref{diff_transfo}, we find
\begin{equation}
        \ddot \K{2} - \beta \K{2} = - \beta g_1^2 \L{2}(\tau_i, \bq) .
\label{eq_order2}
\end{equation}
Thus, the solution is again separable, $\K{2}(\tau, \bq) \equiv g_2(\tau)
\K{2}(\tau_i, \bq)$. The spatial part is given by $\K{2}(\tau_i, \bq) = \L{2}
(\tau_i, \bq)$, and the growth factor is
\begin{equation}
        g_2 = k_a^2 g_{2a} + 2 k_a k_b g_{3b} + k_b^2\, g_{2c} + l_a g_{1a} +
        l_b g_{1b} \ .
\label{eq_order2_full_time}
\end{equation} 
For $\Omega =1$, we simply find $g_{2a} = -\frac{3}{7} g_{1a}^2,\ g_{2b} =
\frac{3}{2} g_{1a} g_{1b},\ g_{2c}^2 = - \frac{1}{4} g_{1b}$, while for $\Omega
< 1$,
\begin{eqnarray}
        g_{2a} &=& 1-{9\over 4}(\tau^2 -1)\left[\tau + (\tau^2-1)S\right]^2 ,
        \nonumber \\ g_{2b} &=& -{3\over 4}(\tau^2 -1)\left[\tau^3 +
        (\tau^2-1)S\right],
\label{sol_order2_om_inf1}
        \\ g_{2c} &=& -{1\over 4}(\tau^2 -1)^3 .\nonumber
\end{eqnarray}

If only a growing mode is initially present ($k_b = 0 = l_b$), which we assume,
one must have $g_2/g_1 \rightarrow 0$ when $\tau \rightarrow \infty$, \ie, when
$\Omega \rightarrow 1$. A limited expansion of
equation\eqref{sol_order2_om_inf1} then shows that $l_a$ must be equal to
$-{3\over 2}$, and this yields the physically relevant solution
\begin{equation}
        g_2 = -{1\over 2} -{9\over 2}(\tau^2 -1)\left\{ 1 + \tau S + {1\over 2}
        \left[ \tau + (\tau^2-1)S\right]^2 \right\} ,
\label{sol2_phys_om_inf1}
\end{equation}
which behaves near $\Omega =1$ as
\begin{equation} 
        g_2 \asymp - {3\over 7}\, \Omega^{-2/63}\, g_1^2 ,
\label{g2_asymp}
\end{equation}
while $g_2/g_1^2 \rightarrow -1/2$ when $\Omega$ approaches $0$. The $\Omega >
1$ solutions are in~\cite{lagrange}. The logarithmic derivative of the second
order growth rate, $f_2 \equiv \frac{a}{g_2}\, \frac{{\rm d} g_2}{{\rm d} a}$,
behaves as $f_2 \asymp 2\,\Omega^{5/9}$ near $\Omega=1$. Alternatively, it can
be somewhat better approximated by $2\,\Omega^{4/7}$ for $0.1\la \Omega\la 1$.

The third order solution is also know and was obtained in \cite{lagrange}. It
is also separable in the $\Omega=1$ case. But when $\Omega \not= 1$, it must be
written as a sum of 2 separable pieces.

\section{APPLICATIONS\label{sec:lag_app}}

Here I show how moments of the PDF can be computed in Lagrangian perturbation
theory, and recover some of the result reviewed above. In particular I give the
skewness for rather general non-gaussian initial conditions. Then I show how
the redshift space distortion effect on skewness can be computed. Finally, I
discuss using the exact perturbative solutions as approximations to the real
dynamics.

\subsection{Moments in the Lagrangian approach}

Since the unsmoothed density contrast is given by $\delta = J^{-1} -1$, we have
\begin{displaymath}
        \VEV{\delta^n( \rx )}_{\rx} = \VEV{(J^{-1} - 1) ^n}_{\rx} =
        \VEV{(J^{-1} - 1)^n\, J}_{\rmq} .
\end{displaymath}
The variance and skewness are then given up to the fourth order by
\begin{displaymath}
\begin{array}{ll}
        \VEV{\delta^2} = &\ \eps{2}\,\VEV{\jjac{1}} + \eps{3}\,\VEV{2
        \jac{1}\jac{2} - \jjjac{1}}+ \\ &\eps{4}\,\VEV{\jjjjac{1} - 3
        \jjac{1}\jac{2} + \jjac{2} +2 \jac{1}\jac{3}} + \ord{5} \\
        \VEV{\delta^3} = & -\eps{3}\,\VEV{\jjjac{1}} + \eps{4}\,\VEV{2
        \jjjjac{1} - 3 \jjac{1}\jac{2} } +\ord{5} ,
\end{array}
\end{displaymath}
where all averages on the displacement field are taken with respect to the
Lagrangian unperturbed coordinate $\rmq$.

Now consider the case of the initial density fields which can be generated by a
displacement field $\bdist{1}$ which three components are independent, with the
same (but arbitrary) statistical law to insure homogeneity and isotropy. While
not fully general, the generated density field $\delta_i = \varepsilon\del{1}$
encompasses a rather wide class of non-gaussian Initial conditions. We note
$\Var$ the variance of any component $i$ of the (linear) gradient field
\begin{displaymath}
        \Var = \VEV{{\Psi}^{(1)\, 2}_{i,i}} = \eps{2}\,g_1^2\,\VEV{\delta_i^2}
        /3 ,
\end{displaymath}
and $\Skew$ its third moment $\Skew = \VEV{{\Psi}^{(1)\, 3}_{i,i}}$, and
$\Kurt$ its reduced fourth moment $\Kurt=\VEV{{\Psi}^{(1)\, 4}_{i,i}}- 3
\VVar$.  We thus have $\VEV{\jjjac{1}} = 3 \Skew$, $\VEV{\jjjjac{1}} = 3 \Kurt
+ 27 \VVar = 3 \Kurt + 3 \VEV{\jjac{1}}^2$. The other term in $\VEV{\delta^3}$
involves the product $\jjac{1}\jac{2}$ which can readily be estimated since
[after\eqref{def_exp_jac} and \eqref{defs}] we have
\begin{displaymath}
        \jac{2} = (1 + g_2/g_1^2)\, \sum_{i>j} ( \dis{1}_{i,i} \dis{1}_{j,j} -
        \dis{1}_{i,j} \dis{1}_{j,i} ) .
\end{displaymath}
It follows by development that $\VEV{\jjac{1}\jac{2}} = 6 (1 + g_2/g_1^2
)\VVar$. We thus have
\begin{eqnarray}
       S_3 &=& \frac{-3 \Skew + \varepsilon g_1 6(\Kurt+3(2-g_2/g_1^2)\VVar)}
        {\varepsilon g_1\VVar(3-3\varepsilon\Skew/\Var)^2}.  \\ &=&
        -\frac{\Skew}{3\VVar g_1 \varepsilon}+ 4-2\, g_2/g_1^2 +
        2\,\frac{\Kurt\Var-\Skew^2}{3 \VVVar} + \cal{O}(\varepsilon).
\label{s3-gen}
\end{eqnarray}
For an initially Gaussian field with $\Skew = \Kurt = 0 $, we get the simple
result
\begin{equation}
        S_3 = 4 - 2\, g_2/g_1^2 + \ord{2} ,
\label{S3R}
\end{equation}
whose first term corresponds to the pure Zeldovich approximation and had been
found by Grinstein and Wise \cite{grin_wise}. The value of the ratio of growing
modes may be obtained from the exact solutions given before, but it is hard to
imagine practical cases when our approximation\eqref{g2_asymp} might not be
sufficient. We thus have the handy and quite accurate formula for a Gaussian
initial field with $0.1 \la \Omega \la 2$ (Bouchet \etal 1992)
\[
        S_3 \approx {28 + 6\,\Omega^{-2/63} \over 7} ,
\]
which generalizes the $S_3 = 34/7$ found by Peebles (1976) in the $\Omega = 1$
case. The more general formula\eqref{s3-gen} gives the time-evolution of the
skewness factor $S_3$ in spatially flat models for the class of non-Gaussian
initial conditions which may be generated by independent displacement fields
along three axes. The general formula for the $\Omega = 1$ case can be found in
\cite{fry-scherrer}.

\subsection{Redshift Space Skewness\label{par-redshift}}

In redshift space, the appearance of structures is distorted by peculiar
velocities. At ``small" scales, this leads to the ``finger of god'' effect: the
clusters are elongated along the line-of-sight due to their internal velocity
dispersion. This is an intrinsically non-linear effect, and we shall not be
concerned with it. At ``large'' scales, the effect is reversed: the coherent
inflow leads to a density contrast increase parallel to the line-of
sight. Indeed, foreground galaxies appear further than they are, while those in
the back look closer, both being apparently closer to the accreting structure
(Sargent \& Turner 1977; LSS, \S76; Kaiser 1987).

Kaiser (1987) estimated this redshift effect, in the large sample limit, on the
direction averaged correlation function, and found $\VEV{\delta^2}^z = (1 +
2/3\,\Omega^{0.6} + 1/5\,\Omega^{1.2})\,\VEV{\delta^2}$, where the superscript
$z$ correspond to a redshift space measurements. For $\Omega =1$,
$\VEV{\delta^2}^z = 28/15 \VEV{\delta^2}$. We now turn to a Lagrangian analysis
of this effect.

\begin{figure}[hbtp] 
\label{fig-s3-z}
\hbox{
\psfig{file=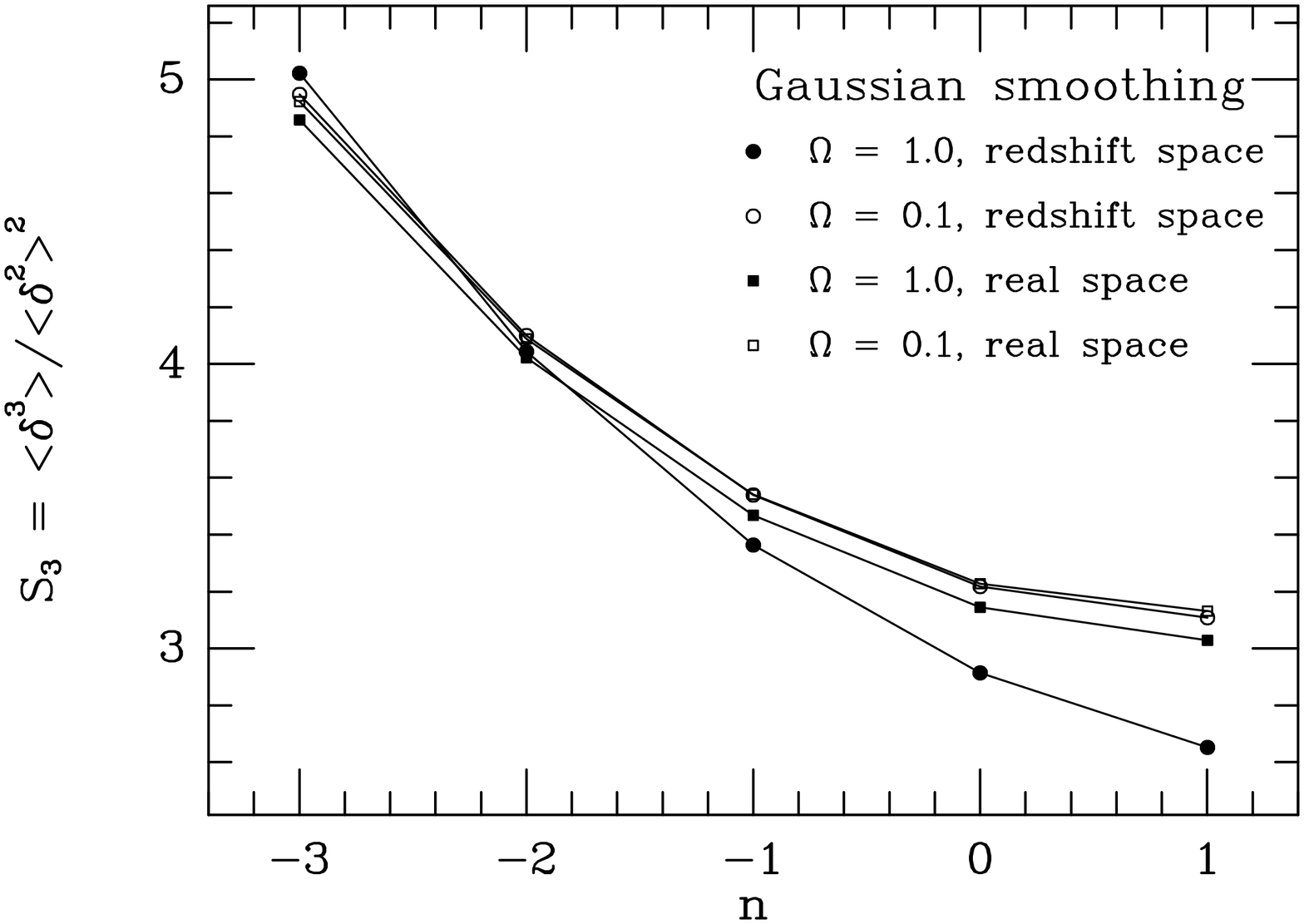,bbllx=58pt,bblly=93pt,bburx=530pt,bbury=762pt,width=0.5\textwidth,angle=-90}
\psfig{file=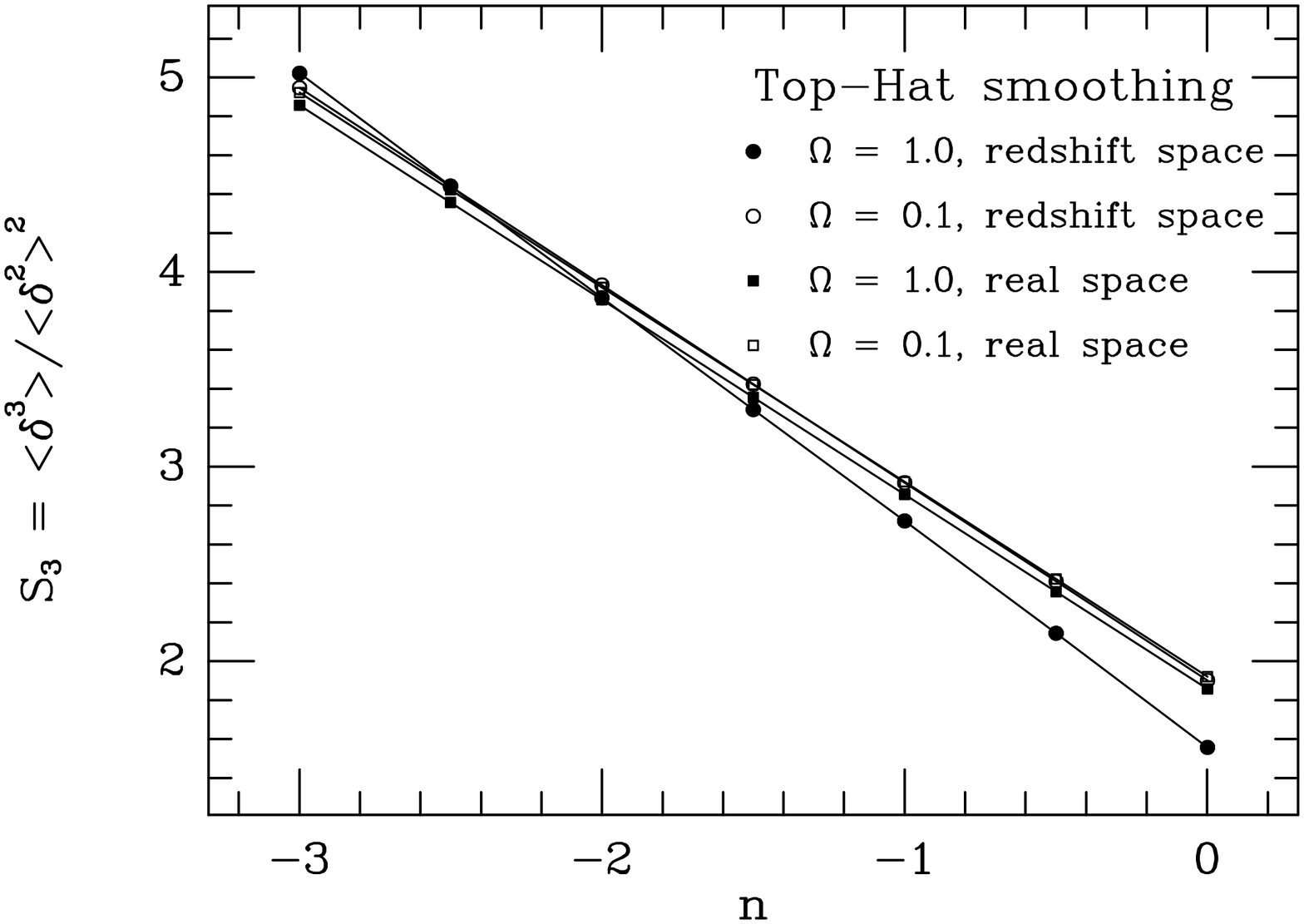,bbllx=58pt,bblly=93pt,bburx=530pt,bbury=762pt,width=0.5\textwidth,angle=-90}
}
\caption{See legend in text. Courtesy Hivon \etal 1993)}
\end{figure}

Let us consider the case of spherical coordinates, when distances to the
observer would be estimated by means of redshift measurements. And let us now
denote redshift space measurements by the superscript $z$. The redshift space
comoving position $\bx^z$ of a particle located in $\br (\bq ) = a \bx (\bq )$
is $\bx^z = \dot\br /(aH)$ (with $H=\dot a/a$). The real space perturbative
expansion\eqref{define_lagrange} is then replaced by
\begin{eqnarray}
        \bx^z = \bq & + & \left[ 1 + f_1(t) \right] g_1(t) \bdist{1}(\bq)
        \nonumber \\ & + & \left[ 1 + f_2(t) \right] g_2(t) \bdist{2}(\bq) +
        \ord{3} ,
\end{eqnarray}
where we have explicitly used the separability of $\bdis{1} = g_1(t)
\bdist{1}(\bq)$ and $\bdis{2}= g_2(t) \bdist{2}(\bq )$.  In the limit of an
infinitely remote observer, say along the $r_3$-axis, we can approximate
spherical coordinates by Cartesian ones. The observed density contrast
$\delta_{z}$ in comoving coordinates is then simply $\delta^{z}( x_1, x_2,
x_3^z)$. All we have to do, thus, is to replace everywhere in the calculation
of $S_3$ the displacement field along the third axis
\begin{displaymath}
        \dis{m}_3=g_m\,\dist{m}_3 \quad{\rm by}\quad (1+ f_m)g_m\,\dist{m}_3
\end{displaymath}
for $m=1$ and $2$. We have for instance, $\VEV{\jjac{1}} =
\left[2+(1+f_1)^2\right]\Var$.

This shows that in our so-called ``infinite observer limit'', we have
$\VEV{\delta^2}^z = (1+2/3\, f_1 + 1/3\, f_1^2) \VEV{\delta^2}$ which slightly
differs from Kaiser's calculation (who did a calculation in spherical
coordinates instead of our rectangular ones). Indeed, we find that the redshift
space variance is boosted by a factor 30/15 for $\Omega =1$, instead of his
value of 28/15.

The calculation of the terms involved of the skewness calculation is fairly
straightforward, with the exception of a piece ${\cal E}$ which is small and
can be bounded. The result for our fairly general non-gaussian model is
\begin{eqnarray}
        S_3^{z} &=& 6-{\left( 2 + (1+f_1)^3 \right) \, \Skew \over
        \varepsilon\,\left( 2 + (1+f_1)^2 \right)^2\, \VVar } \nonumber \\
        &-&2\, {\left( 2 + (1+f_1)^3 \right)^2 \,\Skew^2 \over \left( 2 +
        (1+f_1)^2 \right)^3\,\VVVar } \nonumber \\ &+&2\, {\left( 2 + (1+f_1)^4
        \right) \,\Kurt \over \left( 2 + (1+f_1)^2 \right)^2\, \VVar }
        \nonumber \\ &-& 6\,{ 1+2(1+f_1)^2 +(g_2/g_1^2)\left[ 3 + 2f_1 +f_2
        +f_1f_2{\cal E}\right] \over \left[ 2 + (1 + f_1)^2\right]^2},
\label{S3Z}
\end{eqnarray}
with $ 1\geq{\cal E}\geq 0$. Of course, we recover the real space
result\eqref{S3R} if we set $f_1 = f_2 = 0$. On the other hand, if $\Omega =1$,
we have $f_1 =1 =f_2/2$ (and $g_2/g_1^2 = -3/7$), which yields for Gaussian
initial conditions $S_3 = (35+{\cal E})/7$ while, for $\Omega=0.1$, $S_3
\approx (34.5+0.4{\cal E})/7$ (to be compared with the value of 34/7 in real
space).

The formula\eqref{S3Z} obtained above applies only in the limit of large
volumes (like Kaiser's result), since we have taken the limit of an infinitely
remote observer. A full calculation along the previous lines, but in spherical
coordinates, and including the effect of smoothing, is done in \cite{s3z}. In
any case, equation\eqref{S3Z} clearly shows that the ratio $S_3$ is, for
Gaussian initial conditions, nearly independent of the value of $\Omega$, and
is barely affected by redshift space distortions. It is interesting to note,
though, that in the non-Gaussian case, the distortion might be rather large,
for large enough \Skew\ or \Kurt.

\subsection{Approach of the Non-Linear Regime\label{sec:anl}}

So far, we have mainly considered rigorous uses of the Lagrangian perturbative
approach, for instance the derivation of a second-order quantity, the skewness
of a PDF, with the help of second order perturbation theory.

Now, we examine to what extent the second-order theory brings improvement to
Zeldovich approximation, when both are used as approximations to the real
dynamics, i.e. outside of their rigorous validity range.

The first and second order solutions in Eulerian and Lagrangian perturbation
theory were compared in \cite{lagrange} to spherically symmetric cases whose
evolution is analytically known. For instance, the results of those
approximations were checked for the density and the divergence of the velocity
field in the spherical top-hat case, when its amplitude is varied. Other
comparisons were made also for given smooth profiles.

Of course, direct comparisons with numerical simulations
(e.g. Ref.~\cite{lagrange} or Buchert, this volume, and references therein)
were also performed. It turns out that the Lagrangian approximations perform
considerably better on a given non-linear scale $\ell$ if they are applied to
initial conditions smoothed on that scale. This avoids amplifying the small
scale ``noise'' whose structure cannot be followed anyway.

These comparisons lead to the following approximate ranking (at least for
moderate final density contrasts $\sim 1$):
\begin{displaymath}
\begin{array}{ll} {\bf density\ contrast:} 
        & \\ {\rm Lagrangian\ second\ order}\ & >\ {\rm Zeldovich} \ \simgt \\
        {\rm Eulerian\ second\ order}\ & >\ {\rm Eulerian\ linear\ theory},
\end{array}
\end{displaymath}
\begin{displaymath}
\begin{array}{ll} {\bf Velocity\ field:} &
        \\ {\rm Lagrangian\ second\ order}\ & \sim {\rm Eulerian\ second\
        order} \ > \\ {\rm Zeldovich} \ &\sim {\rm Eulerian\ linear\ theory}.
\end{array}
\end{displaymath}
Here the signs ``$>$'' and ``$\sim$'' mean respectively ``more accurate than''
and ``of comparable accuracy to''.  For relatively large final density
contrasts, the Eulerian approach becomes particularly inefficient, except for
the velocity field, for which it tends however to be less accurate than the
Lagrangian one. The second order Lagrangian approach gives, for moderate final
$\delta$, an excellent approximation of the density contrast and the velocity
field. Its seems to be able to reproduce density contrasts as large as ten.

There is hope that these lagrangian approach to the non-linear regime might
lead us to a better understanding of this most poorly known of all regimes, the
transition between the weakly and strongly non-linear realm. It has been
recently an area of considerable activity and we shall have to wait till the
dust settles.
 
\section{CONCLUSIONS}

Perturbation theory already has a long and distinguished history in Cosmology,
but for many years the attention focused mainly on the linear terms, \eg for
early universe calculations of the power spectra of light and matter at
recombination. But for the latter evolution of large scale structures,
perturbation theory went out of fashion in the era which saw the advent of
massive computer simulations of their formation. These allowed investigating
strongly non-linear scales, the only ones that could be carefully studied
statistically in the galaxy surveys of the time. Another reason of this
purgatory period of the theory may have been the perception that any early
signature of the weakly non-linear phase would have been totally erased by the
following strongly non-linear phase, \ie that it was irrelevant. In any case,
the situation has now dramatically changed, galaxy catalogs encompass ever
increasing scales where density contrasts are weak; they display remarkable
moments hierarchy on large scales, with a striking similarity to that observed
on smaller scales. Meanwhile realistic calculations have become recently
available (\eg including the smoothing inevitable in any realistic measurement,
redshift distortions, etc), and last but not least, comparisons with numerical
simulations showed the predictive power of the theory. In a few years the
beauty awakened, and this area of research is now striving. These notes just
surveyed some of the simpler analyses, and new results are posted nearly every
week. Of course, much remain to be done, and maybe the most important question
to be answered is why does these perturbative approaches work so well?


\newcommand{\inprep}{{\em in preparation}} \newcommand{\inpress}{{\em in
press}} \newcommand{\submit}{{\em submitted to}} \newcommand{\preprint}{{\em
preprint}}

\newcommand{\mn}{{\em Mon. Not. R. astr. Soc}} \newcommand{\apj}{{\em
Astrophys. J.}}  \newcommand{\apjsup}{{\em Astrophys. J. Suppl.}}
\newcommand{\aj}{{\em Astron. J.}}  \renewcommand{\aa}{{\em Astr. Astrophys.}}
\newcommand{\ass}{{\em Astrophys. Space Sci.}}  \newcommand{\nat}{{\em Nature}}
\newcommand{\et}{{\em et al.}\,\,}

\section*{References}

\vfill

\end{document}